# Specific instrumentation and diagnostics for high-intensity hadron beams

*Kay Wittenburg*
DESY, Hamburg, Germany

**Abstract**
An overview of various typical instruments used for high-intensity hadron beams is given. In addition, a few important diagnostic methods are discussed which are quite special for these kinds of beams.

## 1   Introduction

All beam instrumentations for high-intensity hadron beams have to fulfil one important criterion: the instruments have to be as minimally invasive as possible to survive the full beam. If, for any reason, this cannot be achieved, the required diagnostics cannot be done with the full intensity of the beam and interpolations are necessary to calculate the parameters of the nominal beam. Such an interpolation might generate large error bars and therefore might not be suitable for precise beam diagnostics.

A second important feature of the instrumentation is the required dynamic range [1]. Typically the instrument has to cover signals coming from low-intensity beams during commissioning up to very-high-intensity beams after an upgrade of the accelerator (which often does not include an upgrade of the beam instrumentation). Sometimes tiny "pilot bunches" have to be diagnosed to ensure that the whole accelerator chain has been set up correctly before injecting the full beam. Also variable modes of operation, e.g. continuous wave (CW) beams, various ion types, long and short pulsed beams, have to be diagnosed. Often the beam has a large diameter, especially non-relativistic beams. Therefore, large size beam monitors with large apertures are needed.

A third important feature of high-intensity beam instruments is that some diagnostic systems have to create a beam interlock or allow the signal to protect the machine against damage from mis-steered or unmatched beams. Therefore, their high reliability and availability as well as their accurate and stable work are necessary to ensure high productivity of the accelerator.

The following sections summarize the most important instruments for sufficient beam diagnostics of high-intensity hadron beams with an emphasis on minimal invasive devices and their high dynamic range. The chapters are followed by some examples of special beam diagnostics which are important for high intense hadron beams. Instruments mainly used in electron accelerators are not mentioned here (e.g. cavity BPMs (beam position monitors) ICTs (inductive current transformer), synchrotron radiation from bending magnets, etc.). An example for the main instruments in high-intensity accelerators is given in Table 1; it summarizes the various beam diagnostic components of the J-PARC complex.

Table 1: Summary of the beam diagnostic components of the J-PARC complex; from Refs. [2–5]

| LINAC: MEBT, DTL/SDTL, A0BT, L3BT | 103 Beam position monitors (BPMs) |
| --- | --- |
| | 98 Slow and fast current transformers (SCTs/FCTs) |
| | 34 Profile monitors (wire scanners (WSs) and destructive halo monitors (beam scraper monitors (BSMs)) |
| | 125 Beam loss monitors (BLMs; scintillators and proportional chambers) |
| RCS | 62 BPMs |
| | 9 Current monitors (direct current transformer (DCCTs), SCTs, FCTs, wall current monitor (WCMs)), WCMs used for bunch length measurement. |
| | 7 Secondary emission monitors (SEMs) |
| | 2 Ionization profile monitors (IPMs), also for halo monitoring |
| | 134 BLMs (scintillators, proportional chambers, ionization chambers) |
| Beam transfer lines: 3–50 BT 3 NBT | 19 BPMs |
| | 5 FCTs |
| | 5 SEMs |
| | 53 BLMs (Proportional and ionization chambers) |
| Main ring (MR) | 192 BPMs |
| | 11 Current monitors (DCCTs, FCTs, WCMs), WCMs used for bunch length measurements. |
| | 238 BLMs (proportional and ionization chambers) |
| | 6 Screen monitors (SEMs, luminescence screens) |
| | 3 Profile monitors (WSs, IPMs) |

## 2 Instruments for beam current and position measurements

Typically the electromagnetic field of the particle beam is used to determine its charge (current) and position. Its signal spectrum extends from the DC component of the beam to its radio-frequency (RF) frequency (neglecting bunch sub-structures). All electric and most magnetic signals cannot reach the region outside the conducting and non-magnetic beam chamber due to its effective shielding. Only the magnetic DC component of the beam can be detected outside the chamber while the much more useful part of the spectrum lies at higher frequencies and is therefore only accessible inside the chamber or through a "gap" in the beam pipe. A thin ceramic ring soldered at both ends of the beam pipe is required to form such a non-conducting gap. Typical beam current monitors make use of such a gap while BPMs use antennas inside the chamber together with a small ultra-high-vacuum (UHV) feedthrough to gain access to the signal at the outside of the chamber.

### 2.1 Resistive wall current monitor

The interruption in the beam pipe by a gap forces the image current to find a new path. By clever design of the monitor, the path and its impedance $Z_{gap}$ are defined by the instrument designer. The voltage across the gap $V_{gap}$ is then

$$V_{gap}(\omega) = Z_{gap}(\omega) \cdot I_{beam}(t) = Z_{gap}(\omega) \cdot (-I_{wall}(t))$$

with

$$\frac{1}{Z_{gap}} = \frac{1}{R} + \frac{1}{i\omega L} + i\omega C$$

The resistance $R$ is formed by a resistive network across the gap, the inductance $L$ and the capacitance $C$ depend on geometrical and mechanical issues. Here $Z_{gap}$ is typically of the order of a few Ohms. Many detailed design studies are necessary to achieve a flat response of the monitor over a large frequency range and to avoid dependence from the beam position:

- An UHV compatible ceramic (with relative permittivity of around 10) forming an equal spacing of the gap and a separation from the beam vacuum. The gap should be short compared with the bunch length to avoid beam shape integration. Since $C$ depends on the gap width, it should not be too small to avoid high $C$ and therefore a limit in the bandwidth. Typical values for $C$ are about tens of picofarads.
- Equally spaced resistors of the same value $R^*$ around a round gap and signal summation by combining the signals from four quadrants avoiding beam position dependence.
- Well-defined bypass for image current and avoiding resonances by adding material with high µ. This increases the inductance $L$ at low frequencies and reduces the lower cutoff frequency. Typical values for $L$ are around 100 nH.
- Reducing other, stray currents vagabonding along the pipe by careful shielding and grounding.
- Avoiding monitor positions close to beam pipe discontinuities, since higher-order modes above cutoff can travel significant distances in the beam pipe. Even some absorbing material (e.g. ferrites) on both sides of the gap but inside the beam pipe might be useful to reduce high-frequency background [6].

Many useful design hints are given in Refs. [7, 8]. A sketch of a WCM is shown in Fig. 1. This type of monitor can have a broad frequency response from a few kilohertz up to a few gigahertz[1] with flat impedance. A frequency response variation of less than 1 dB over the full range was reached [7]. The low-frequency cutoff leads to a droop in successive signals, which has to be taken into account.

The bunch current or the number of particles in the bunch $N_B$ can be determined by

$$N_B = \frac{\int V_{gap} \cdot dt}{Z_{gap} \cdot e \cdot K}$$

while $K$ is a constant which takes into account various attenuation of cables, combiners and calibration constants. Note that the wall current does not contain information about the DC current component of the beam since this frequency component of the beam current penetrates the (non-magnetic) beam pipe unaffected [10]. Therefore, the baseline of the signal is shifted while the shift is proportional to the DC current. A careful baseline restoration is needed for precise bunch current measurements.

Owing to its broad frequency response the wall current monitor (WCM) is often used, in addition to bunch current determination, for measuring the longitudinal profile of the bunch, calculating its emittance and diagnosing longitudinal instabilities. Note that the ultimate bandwidth is limited by the spreading angle of the radial electrical field lines which have an opening angle of approximately $1/\gamma$. This limits the longitudinal resolution for non-relativistic beams. WCMs are also used for RF and timing feedback issues (e.g. compensating beam loading) and time-of-flight (TOF; energy) measurements since they provide very fast and large signals.

Some care has to be taken at high beam currents:
The absorption of higher-order modes (HOMs) in the magnetic material will increase its temperature. A good cooling is necessary. This is especially true for short bunches as in electron accelerators which induce strong HOMs.
High beam currents may cause saturation in the magnetic material which changes the inductance $L$ and therefore the lower cutoff frequency. As a result the droop rate will change.

---

[1] A recently developed wall current monitor for CLIC reached a bandwidth of 20 GHz [9].

The power level in the monitor resistors $R^*$ can reach some tens of Watts at $N_B = 10^{11}$ particles/bunch which may lead to high thermal load of the resistors at high repetitive signals. Many distributed resistors around the gap will help but a change in resistance with temperature might occur which will change the calibration constant of the monitor. The signal level may reach 100 V or more. Detailed calculations of the power levels for the WCM in the Large Hadron Collider (LHC) can be found in Ref. [11].

Since the signal level is quite high, a high dynamic readout can be achieved by various methods like switch able attenuators/amplifiers, logarithmic amplifiers or large bit analogue-to-digital converters (ADCs). The last two options are limited in bandwidth, so that a compromise has to be found for each specific application.

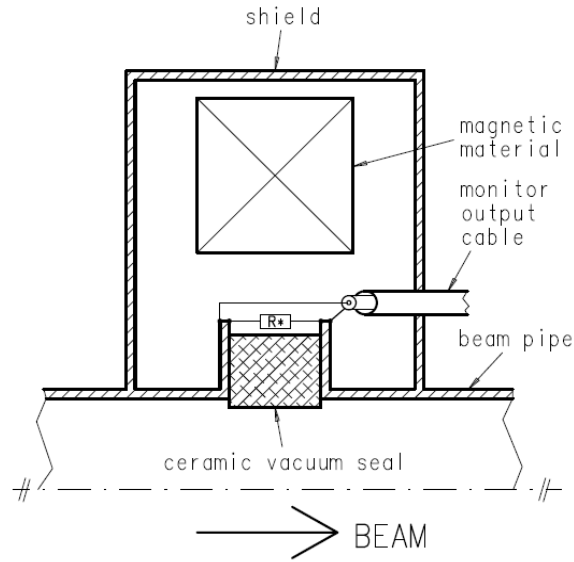

**Fig. 1:** Conceptual view of a WCM. The resistors $R^*$ are distributed evenly around the gap. A magnetic material with high μ is also shown as one coaxial cable for the readout. (Courtesy of M. Siemens, DESY.)

### 2.2 Inductive alternating current transformers

In contrast to the capacitive (electrical) coupling of a WCM, the inductive current transformers are using the magnetic field of the beam to determine the beam current. A bunch crossing a (ceramic) gap in a beam pipe induces a magnetic flux in a high-permeability toroid around the gap, like a primary single turn winding in a classical transformer. It induces a current in a secondary winding of $N_s$ turns and an inductance $L_s$. This current can be measured by the voltage $V_s$ across a resistor $R_s$. By applying the classical transformer equations one receives the typical transformer response [10, 12]:

$$V_s(\omega) = \frac{i \cdot \omega \cdot \tau}{1 + i \cdot \omega \cdot \tau} \cdot R_s \cdot \frac{I_{Beam}}{N_s}$$

with $\tau = L_s/R_s$. Here $\tau$ is the time constant of the secondary winding. For $\omega \gg 1/\tau$ it results in a simple proportionality of

$$V_s = R_s \cdot \frac{I_{Beam}}{N_s}$$

where the measured voltage $V_s$ is proportionally to the beam current $I_{Beam}$ and in phase with it.

The inductance $L_s$ depends on the permeability µ of the core material, the number of windings $N_s^2$ and its dimensions. Assuming typical values for $L_s$ = 1 mH and a load resistor $R_s$ of 50 Ω one obtains the proportionality above for frequencies ω >> 50 kHz. The high-frequency performance of such a classical current transformer is limited by stray capacitance between the windings and to ground as well as due to energy loss in the toroid material (~$ω^2$). Typically the upper limit is in the some hundreds of megahertz range. Proper impedance matching and low-pass filters are essential to avoid resonances at higher frequencies in the readout loop. These limits make this simple AC transformer not suitable for the measurement of the longitudinal bunch shape but it is widely used for bunch charge/current measurements.

Since this device is a classical transformer, it cannot transmit the DC component. Therefore a certain droop rate is indispensable ($τ_{droop}$ ~ $L_s/R_s$); see Fig. 2. An optimization between the high-frequency response and low droop rate becomes necessary. Fast current transformers with a droop rate of <1 %/µs and an upper frequency of more than 800 MHz are commercially available [13]. Accurate DC baseline restoring is necessary, however, to avoid a measurement error in a train of successive bunches. In an accelerator/storage ring with an infinitely long bunch train an equilibrium is reached when the area below and above the zero line are equal.

An advantage of an inductive current transformer is its small dependence on the beam position. Careful magnetic shielding of the core is very important as well as a good shielding of the signal windings to avoid contamination of external noise sources. An absolute calibration of the measured value can be done by simply adding a calibration winding around the core. The response of a well-defined short calibration pulse can be used to calibrate the device. Even drifts can be compensated for by sampling of the calibration signal just before or after the passage of the beam.

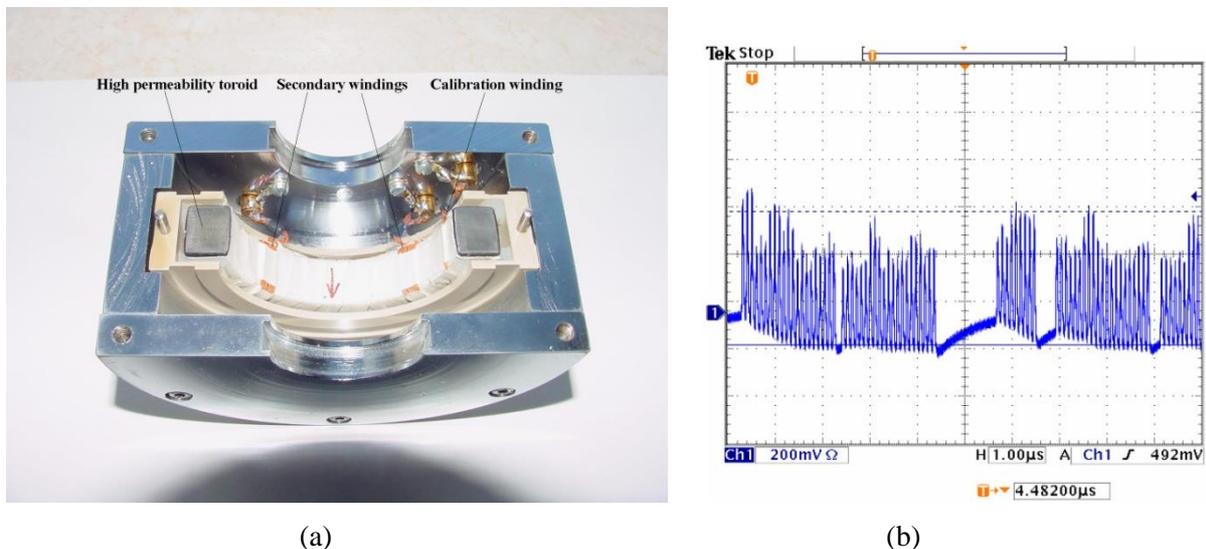

(a)            (b)

**Fig. 2:** (a) An open inductive current transformer at DESY. The magnetic core (toroid) is split into two half to allow easy mounting around a ceramic gap without opening the vacuum. (Photo by N. Wentowski, DESY.) (b) Bunch trains in HERA measured by an inductive AC current monitor. Note the droop and recovery of the baseline in the presence of signals and in the bunch gap, respectively.

High peak currents can cause magnetic core saturation which might result in non-linear behaviour. Therefore, the choice of the core material and the design of the monitor have to fit the required bunch charge range. A dynamic range of ≈$10^3$ and a resolution of $10^{-4}$ of full scale can typically be reached which is quite sufficient for measuring the variation in the bunch charge. Since the voltage output is proportional to the bunch charge only the peak voltage is of interest. The acquisition rate is the bunch repetition rate; maybe twice the rate to obtain a value between two bunches for baseline restoration. High dynamic range (12–14 bit) and high bandwidth ADCs are commercially available with sampling rates up to 100 MHz, which are in most cases sufficient for the required resolution and dynamic range. In circular machines the resolution can be improved by averaging the acquired bunch current over many turns, but taking into account the lifetime of the beam.

## 2.3 Direct current transformers

The integration of any alternating current transformer (ACT) monitor signal over an infinite period is always zero. Precise active baseline restoration may be used to get a DC value of the beam but due to the fact that the baseline slope and level depends on previous beam bunches (several transformer time constants), a precise measurement is difficult. Direct current transformers (DCTs) are used to measure the DC component of a bunched or unbunched beam with high precision and with a dynamic range of $>10^6$. The high dynamic range is required due to the fact that a commissioning of a circular accelerator might be done with a pilot bunch only while the design allows some orders of magnitude more bunches. Sensitivities as low as 0.5 µA exist [13] which is sufficient for low current commissioning. Obviously a DC beam current measurement does not make sense in short pulsed machines such as linacs (except CW linacs) or transport lines, but it is the only device that can measure the beam current of an unbunched or coasted beam in a circular machine.

The principle of a DCT (also called a DCCT, PCT or zero-flux current transformer) relies on a pair of identical toroids with high permeability. They are excited in an opposite direction into saturation by a common AC current (or voltage) into individual windings. Careful matching of the toroids and the exciting current is necessary. A common sense winding picks up the resulting induced current which is zero in the case of a perfect matching. A charged beam crossing the two coils drives one of the two out of saturation which leads to a modulated current in the sense of winding with a frequency of the second harmonic of the exciting frequency. This current is then proportional to the DC beam current. It can either be measured by synchronous detection or more often by a feedback loop which prevents any magnetic flux change in the cores [14, 15]. This increases the useful dynamic range to more than six decades and reduces the recovery time of the device. The bandwidth of such a DCT is limited from DC to some tens of kilohertz. A further reduction in bandwidth is often useful to reduce the low-frequency noise and to extend the resolution. If even more dynamic range is needed the only (costly) solution is then to use two DCTs with different ranges.

Some issues of DCTs are addressed in the following [16]:

- Higher harmonics in the output lead to ripple which needs to be suppressed [17].
- Temperature drifts induce a drift of the baseline. Good temperature stabilization and/or a frequent measurement of the offset in the absence of the beam are recommended.
- HOMs in the gap may induce heating; therefore, water cooling might be necessary. Care has to be taken during vacuum bake-out not to exceed a core temperature above about 60 °C to avoid damage to the core.
- A calibration winding enables an absolute determination of the beam current.
- A DCT is quite sensitive to external noise and especially to external magnetic fields. Therefore a good electrical and magnetic shielding is essential. The magnetic shield should extend along the vacuum chamber with a length of at least twice the diameter of the beam pipe and without any gaps. It should have the highest possible µ, but should not saturate. It is also necessary to shunt all external currents away from the monitor to enable it to measure the beam current only.

Application issues include the following:

- Note that the DC component of a non-relativistic beam in a circular accelerator increases with the real acceleration of the beam particles while the bunch charge remains constant.

- Beam lifetime determination in storage rings is often done by high-precision DCTs [18, 19].
- A method to determine the amount of coasting beam in a storage ring is provided by a comparison of the ACT and DCT monitors. In the absence of a coasting beam (e.g. during or immediately after acceleration) the sum of all individual bunch currents should be equal to the DC component of the beam. In fact, this can be used to calibrate the monitors. An increasing difference between the two monitors indicates an increase of coasting beam in the machine [20].

## 2.4 Bunch shape monitor

The longitudinal charge distribution of a highly relativistic bunch (sometimes called a "bunch shape" or often a "bunch length") can be determined by a high-bandwidth WCM (see Section 2.1). For low-$\beta$ beams the electromagnetic field is not purely transversal and hence does not represent the charge distribution. Hence, a bunch shape monitor (BSM) based on secondary emission is more adequate to image the real distribution in this case. It was originally developed in Ref. [21], developed further in Ref. [22] and is now used in many proton, H$^-$ and ion linacs. The monitor based on a coherent transformation of the temporal structure of the bunch into one of the secondary electrons and then into their spatial distribution. Figure 3 shows its principle: parts of the beam hit a metal wire target (typically tungsten) in the beam pipe (see "1" in Fig. 3). The wire emits low-energy secondary electrons of a few electronvolts. Since this process does not have a significant delay, the temporal structure of the electrons now represents that of the bunch. The electrons are accelerated radially away by a negative bias voltage ($U_{targ} \approx -10$ kV) on the wire. A fraction of the electrons passes a collimator (see "2" in Fig. 3), an electrostatic lens and a varying RF field of a deflector (see "3" in Fig. 3; $U_{RF} = A \cdot \cos(n\omega t + \Phi)$). The RF is a multiple $n$ of the acceleration frequency of the beam and is synchronized in time. The transit time of the electron bunch should be somewhat shorter than half of the wavelength of the RF. Depending on the arrival of the electrons with respect to the phase $\Phi$, the electrons received transversal kick so that their longitudinal distribution is transformed into a spatial distribution after some distance (see "4" in Fig. 3). At that point the distribution can be measured by various detectors, e.g.:

- phosphor screen + CCD [23];
- multichannel electron detector [24];
- scanning phase + stable slit (or stable phase + scanning slit) + collector [25].

In the special case of a H$^-$ beam, the detached electrons originated by dissociation of the H$^-$ ions on the target wire (of initial energy of some kiloelectronvolts) contribute to some background [26]. Energy separation by an additional spectrometer behind the second slit (see "5" reduces the background down to better than $10^{-5}$ of the maximum. This enables measurements of longitudinal halo with a dynamic range of $10^5$ [27]; see also Fig. 4.

The resolution of a BSM is defined by some factors [28, 29]:
(1) the time uncertainty of the secondary electron emission process is far below some picoseconds;
(2) the velocity and angular spread of the secondary electrons is minimized by a high bias voltage of $-10$ kV on the target and a short distance to the first collimator ($\approx 1$ ps);
(3) the RF deflector might need a sufficient strength of up to some 1000 V/cm to get an image on the detector with a resolution of $\approx 1$ ps;
(4) the slit size of the first collimator (the spot size on the detector without RF) should be small compared with the transversal dimension on the detector; this effect can be measured and subtracted;
(5) the phase stability of the RF and of the synchronization can be kept much below 1° of the RF phase;
(6) non-linear effects such as space charge and lens aberrations are assumed to be negligible.

Typical resolution achieved so far lies in the order of some 10 ps.

A BSM for measuring all three dimensions of the bunch is used in Ref. [24]. Here additional the target wire is scanned across the beam and the slit of the first collimator is scanned along the wire. The intensity distributions versus the scanning positions give the two transversal distributions of the beam. A translation of the whole BSM along the beam axis enables the measurement of a phase difference ΔΦ between the two locations and therefore a velocity measurement of the beam particles becomes possible [30]. A translation distance of a few centimetres is sufficient to determine velocities of up to $\beta \approx 10\%$.

The energy deposition of the beam in the target implies two problems: first the creation of thermal electrons and second the melting of the target. Thermal electrons blind the whole monitor. Therefore, care has to be taken in positioning the target into the beam centre: an off-centre measurement increases the lifetime of the target. To overcome this problem the electrons created by the ionization of the residual gas are used for high-intensity beams [31]. Since the electrons have a broad spectrum of energies an electrostatic analyser is located just after the first collimator to ensure a mono-energetic beam at the RF deflector.

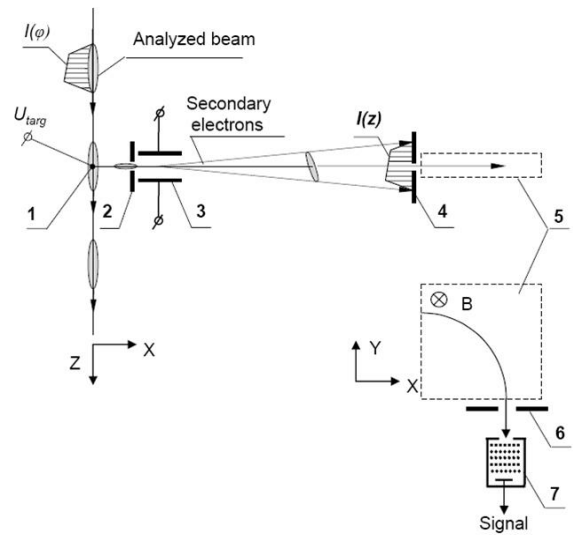

**Fig. 3:** Configuration of a BSM: 1, target wire; 2, input collimator; 3, RF deflector combined with an electrostatic lens; 4, output collimator or screen; 5, detector; consists here of a bending magnet; 6, collimator; 7, secondary electron multiplier (reproduced from Ref. [25])

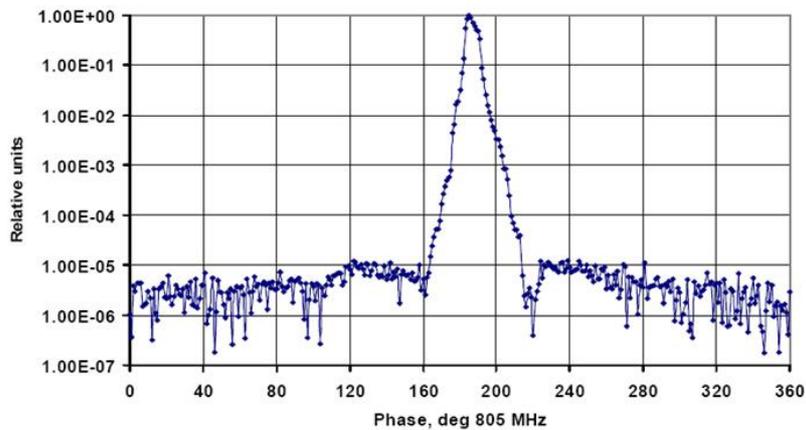

**Fig. 4:** Longitudinal bunch shape at SNS measured by a BSM with a high dynamic range (reproduced from Ref. [25])

## 2.5 Faraday cup

A Faraday cup (FC) is typically a fully destructive device which can be driven into the beam and in which the beam is completely absorbed. A full absorption of the beam enables an absolute determination of the beam charge; therefore, a FC is often used to re- and cross-calibrate the non-destructive current monitors [32]. Since the FC has to collect the whole charge of the beam, no charges must escape the cup:

- It has to be large enough to avoid any leakage of shower and multiple scattered particles. Therefore, its use is restricted to low-energy beams, otherwise its dimension became very large [32]. Hadron beam energy should be kept below about 150 MeV to stay below the π-production threshold.
- Backscattered particles have to be collected.
- Secondary and thermal electrons should not escape the FC.

The last two subjects are solved by additional negative voltage on a repeller electrode of approximately −100 V at the entrance of the cup, sometimes in combination with a magnetic field perpendicular to the incoming beam [33].

FCs (like most intercepting devices) in high-intensity beams need water cooling of the collector to take away the beam power of $P_{Beam} = E_{Beam} \times I_{Beam}$. Note that the water cooling only dissipates the average beam power. For pulsed beams the highest temperature is reached close to the penetration depth of the impact. This drives the design of the electrodes. Special shaped electrodes are necessary for intense low-energy ion beams. These ions have a very short penetration depth in materials which lead to a very dense energy deposition. Therefore a V- or saw-tooth-shaped electrode with a large inclination against the beam axis is needed to distribute the impact over a larger surface [34].

A good isolation of several gigaohms of the electrode offers a large dynamic range of beam charge measurements. Care has to be taken not to deteriorate the isolation (and the de-ionized water of the cooling) by radiation. A good vacuum is essential in the FC to avoid extra accumulated charges due to ionization for the residual gas molecules. An absolute accuracy of better than 1 % can be reached.

A high-bandwidth FC enables a measurement of the longitudinal bunch shape also for low-β beams. A careful design of the collecting electrode is necessary to achieve sufficient bandwidth. A bandwidth of some gigahertz was measured with coaxial and stripline types of electrodes in Ref. [35]. In the case of low-β beams the advanced electrical field of the bunch has to be considered. A grid in front of the electrode is required to shield the cup from this effect [36].

## 2.6 Beam position monitors

The fundamental principle of a BPM is to measure the transversal centre of the electromagnetic field of the beam with respect to the vacuum chamber wall. There are two ways of coupling on the electromagnetic field, by capacitive pick-ups and by inductive pick-ups. Inductive BPMs consist typically of thin loops with their open area parallel to the beam so that the magnetic field of the beam couples into the loop and induces a current (see e.g. [37]). The coupling to the beam is inductive for a thin loop and capacitive for a wide one. Nearly all modern hadron accelerators are using nowadays capacitive pick-ups which are therefore discussed in more detail following the discussions in Refs. [38, 39]. To obtain sufficient information on the beam position, the difference of the field amplitude in up–down and left–right orientation have to be measured by the pick-ups and analysed by the readout electronics.

The electromagnetic field of the beam induces an image charge on a metallic plate which is inside the vacuum chamber and insulated. For the following discussion only one plate is considered, but it is true for all four plates of a BPM. Assuming a bunched beam, the image current $I_{im}(t)$ is driven by the beam charge $Q(t)$:

$$I_{im}(t) = \frac{A}{2\pi dl} \cdot \frac{dQ(t)}{dt}$$

with $A = \pi r^2$ is the area of the electrode, $d$ its distance to the beam centre and $l$ its length. Here $dQ/dt$ depends on the beam current $I_{beam}(t)$:

$$\frac{dQ(t)}{dt} = \frac{l}{\beta c} \frac{dI_{beam}(t)}{dt} = \frac{l}{\beta c} \cdot i\omega I_{beam}(t) \quad \text{with } I_{beam}(t) = I_0 e^{i\omega t}$$

where $\beta$ is the beam velocity. To calculate the voltage drop across a resistor $R$, the capacity $C$ between the plate and the grounded vacuum chamber has to be taken into account, therefore the impedance $Z_{plate}$ of the plate is

$$Z_{plate} = \frac{R}{1 + Ri\omega C}$$

and the voltage $U$ becomes

$$U_{im}(\omega, t) = Z_{plate}(\omega) \cdot I_{im}(t) = Z_t(\beta, \omega) \cdot I_{beam}(t)$$

while $Z_t$ is the "transfer impedance" derived from the above:

$$Z_t(\omega) = \frac{A}{2\pi d} \cdot \frac{i\omega}{\beta c} \cdot \frac{R}{1 + Ri\omega C}$$

This impedance has a high pass characteristic which is shown in Fig. 5. Since high-frequency signals are typically transported via 50 Ω coaxial cables, a low input impedance of the first amplifier is often used. The frequency ω depends on the bunch length, which is respectively dependent on the RF frequency of the accelerator. Therefore, the coupling of a capacitive pick-up to the long bunches (e.g. in the beginning of a hadron accelerator chain) is very week. For frequencies $\omega \ll \omega/RC = \omega_{cut}$ the measured voltage $U_{im}(t)$ becomes

$$U_{im}(t) = \frac{R}{\beta c} \cdot \frac{A}{2\pi d} \cdot \frac{dI_{beam}(t)}{dt}$$

where $dI_{beam}(t)/dt = i\omega I_{beam}$ is the derivative of the bunch length. For $\omega \gg \omega/RC = \omega_{cut}$ the voltage $U_{im}$ follows the bunch length by

$$U_{im}(t) = \frac{1}{\beta c} \cdot \frac{1}{C} \cdot \frac{A}{2\pi d} \cdot I_{beam}(t)$$

In practice, this means that button-type pick-ups with about 100 mm² < $A$ < 500 mm² are used only in high-energy hadron accelerators, since their coupling to the beam is strong due to the short bunch length and the high β (see Ref. [40]).[2]

For low-energy beams, large bunch length and sometimes large apertures $d$, the common way for a sufficient signal is to increase the size $A$ of the pick-up and to use high-impedance amplifiers in the readout. An example is a shoe-box type of BPM which is shown in Fig. 6. It has a large aperture but also large size electrodes which are separated diagonally with respect to the beam. Therefore, the induced voltage on both plates is proportional to the length of the beam projection on the electrodes. These types of BPMs are very linear over nearly their whole aperture [41]. Since they are used for long bunches a high-impedance readout at low frequency can be performed to obtain a useful readout voltage.

---

[2] The same argument applies to nearly all electron accelerators as well.

Stripline types of pick-ups are used in the case where the bunch length is shorter or about the length of the electrode. The electrode of length $l$ forms a wide loop or transmission line between the electrode and the wall of the vacuum chamber. A signal is created by the beam on each end of the line which depends on the characteristic impedance $Z_{strip}$ of the electrode, often $Z_{strip}$ = 50 Ω. Depending on the termination $R$ of the downstream port the signal there is cancelled ($R = Z_{strip}$) or appears partially ($R \neq Z_{strip}$). A complete cancellation at the downstream port happens only if the speed of the beam is equal to the speed of the signal in the transmission line which is almost true for $\beta \approx 1$. In this case a stripline is known as a "directional coupler" since the signal on one port depends on the beam direction. Such a BPM can be used to separate the beam positions of two counter-rotating beams in the same beam pipe [42]. The upstream port always includes the induced signal and the reflected inverted signal separated in time by $\Delta t = 2 \cdot l/c$ (for $\beta = 1$). The characteristic frequency of a stripline signal is defined by $\Delta t$ and is $\omega_{strip} = 2 \cdot \pi \cdot 2 \cdot c/2l$. Assuming a short bunch, the Fourier transformation of the response is the transfer impedance $Z_t(\omega)$ [38]:

$$Z_t(\omega) = Z_{Strip} \cdot \frac{\alpha}{2\pi} \cdot \sin(\omega l / c) \cdot e^{i(\pi/2 - \omega l / c)}$$

where $\alpha$ is the azimuthal coverage angle (width of the electrode). Here $Z_t(\omega)$ shows a maximum response at a bunch repetition frequency of $\omega = \omega_{RF} = n \cdot \omega_{strip}/4$ and zero response at $\omega = \omega_{RF} = n \cdot \omega_{strip}/2$; $n = 1, 2, 3, \ldots$ (see Fig. 7). Therefore the optimum length of the stripline electrode is $l = \lambda_{RF}/4$. If, on the other hand, the bunch length exceeds the length of the electrode, partial cancellation occurs at the upstream port which reduces the signal. This is also valid for $\beta \ll 1$, since the field of the bunch is no longer a pure TEM wave. Note that therefore also the coupling to the four pick-up electrodes became strongly non-linear with the beam position [43, 44].

The beam position of one plane is derived from the difference of the signal of the two opposite electrodes[3] and after applying corrections to the non-linear position response of the BPM. The signals are also intensity dependent; therefore, normalization to the intensity is always necessary. There are various electronic concepts in use which are discussed in detail in Ref. [45]. Their sensitivity has to be as low as the minimum expected bunch charge to produce position readings with the required accuracy and resolution. Their dynamic range is defined by the various intensity conditions, e.g. acceleration of proton and ion beams. A dynamic range of several decades might be necessary in such a case. A bunch-by-bunch measurement of the beam position requires a broadband signal processing. In this case the dynamic range is defined by the variation in bunch charge only (plus position and β variations). An orbit measurement requires a narrowband signal processing with a maximum update rate of the revolution frequency. In this case the required dynamic range is defined by the stored or accelerated current which might also include a variable number of bunches, but due to the lower speed of the signal processing there are electronic components (e.g. ADCs) available which have a very high dynamic range.

A high bandwidth offers some advantages, e.g. a measurement of the individual bunch current by summing up all four signals from the electrodes (to be position independent). In particular, during commissioning of an accelerator this feature becomes important in detecting the positions of beam losses. For very high bandwidth readout the bunch length can be resolved by a BPM. At even higher bandwidth a position variation along the bunch can be observed (head–tail modes) [46].

---

[3] Assuming orthogonal electrode arrangement, in electron accelerators with synchrotron radiation the arrangement might be different to avoid direct irradiation of the plates.

The BPM system resolves the beam position along the accelerator (= orbit). The BPMs are best located at local maxima of the β function (close to focusing quadrupoles in both planes) to obtain maximum sensitivity. This already implies two beam position measurements per betatron wavelength with a four electrode BPM. This is typically sufficient for a useful orbit distortion measurement. Often some BPMs are added at critical regions such as high dispersion, injection, extraction or collimation. In high-intensity beams the beam position excursions have to be carefully monitored and controlled. Too high excursions should cause an alarm signal to the machine protection system to avoid damage to components due to a mis-steered high-intensity beam.

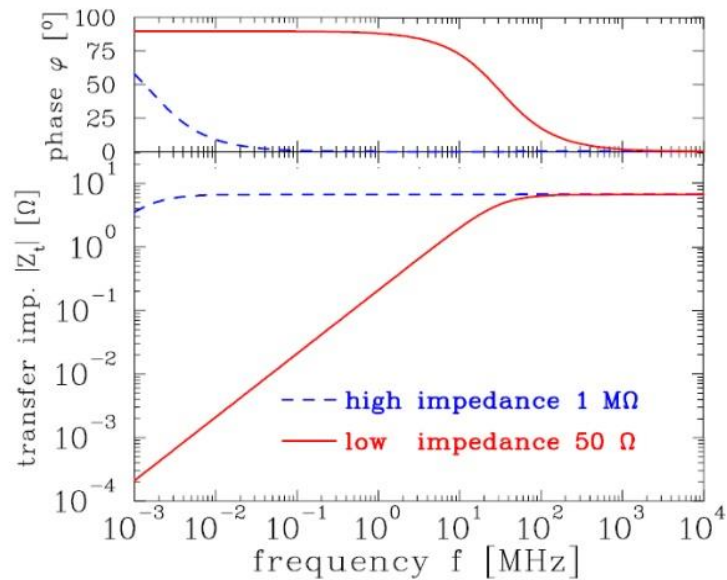

**Fig. 5:** Absolute value and phase of the transfer impedance for a $l = 10$ cm long round pick-up with a capacitance of $C = 100$ pF and an ion velocity of $\beta = 50$ % for high (1 MΩ) and low (50 Ω) input impedance of the amplifier. The parameters are typical for a proton/heavy ion synchrotron. (Courtesy of P. Forck [33].)

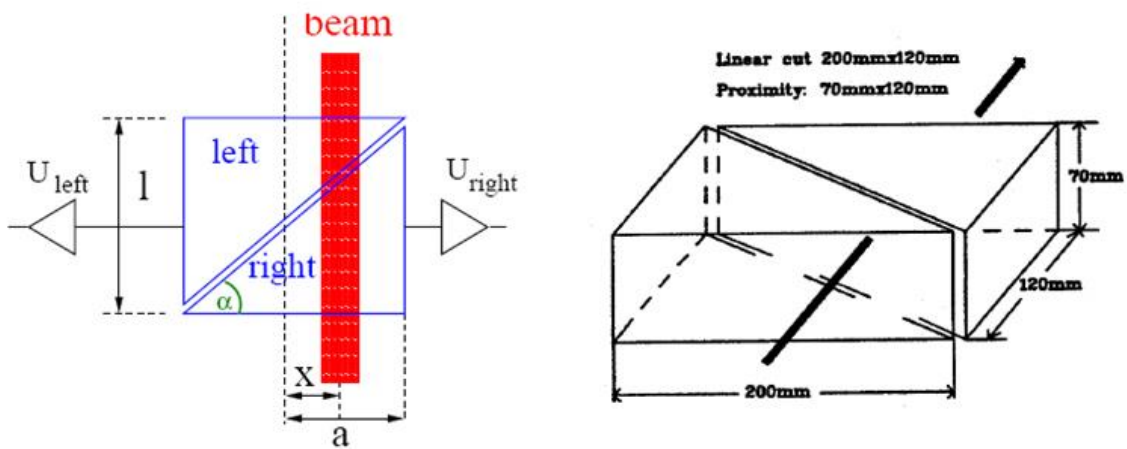

**Fig. 6:** Scheme of the position measurement using a shoebox BPM with linear cut and an example of an electrode arrangement for the horizontal plane (courtesy of P. Forck [33])

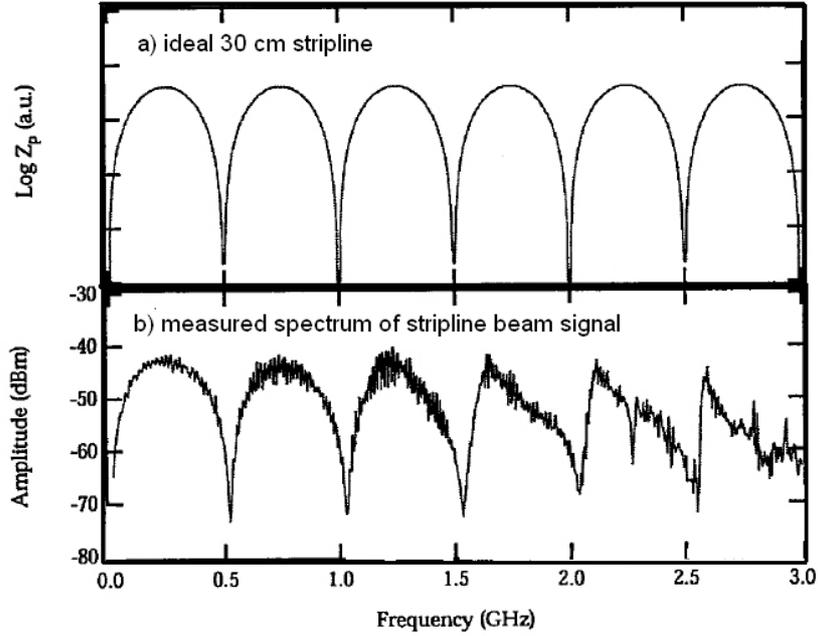

**Fig. 7:** Calculated transfer impedance of (a) an ideal 30 cm stripline and (b) a measurement with a spectrum analyser for a single bunch signal [47].

## 3    Measurement of beam emittance

The transversal emittance ε of a particle beam is used to describe the size of the beam in $x$ and $y$ direction as well as the angular distribution of the particles in the beam in $x'$ and $y'$ directions. The emittance is an ellipse in the $x, x'$ or $y, y'$ plane[4] and its equation can be written as

$$\varepsilon_x = \gamma(s)x^2 \cdot 2\alpha(s)xx' \cdot \beta(s)x'^2$$

where $\alpha(s)$, $\beta(s)$ and $\gamma(s)$ are the Twiss parameters of the respective plane at the position $s$ and $\beta \cdot \gamma - \alpha = 1$. The emittance for Gaussian beam distributions can be expressed by the "Sigma matrix" σ:

$$\sigma = \begin{pmatrix} \sigma_{11} & \sigma_{12} \\ \sigma_{21} & \sigma_{22} \end{pmatrix} = \varepsilon \cdot \begin{pmatrix} \beta & -\alpha \\ -\alpha & \gamma \end{pmatrix} \quad \text{with} \quad \varepsilon = \sqrt{\det \sigma} = \sqrt{\sigma_{11}\sigma_{22} - \sigma_{12}^2}$$

In the geometrical interpretation of the emittance ellipse 39% of the particles lie in the area of π·ε, therefore the typical unit of the emittance is written in π mm mrad. The root mean square (rms) emittance of any phase distribution in $x, x'$ can be calculated by

$$\varepsilon_{rms}(x, x') = \sqrt{\langle x^2 \rangle \langle x'^2 \rangle - \langle xx' \rangle^2}.$$

The matrix element $\sigma_{11}$ is related to the beam size $\sigma_x(s)$ by

$$\sigma_{11} = \sigma_x^2(s) = \varepsilon_x \cdot \beta_x(s).$$

---

[4] In the following $x, x'$ is used for the $x$ or $y$ phase space.

Therefore, the emittance of a particle beam can be determined by measuring the beam size $\sigma_x$, but note that the dispersion $D(s)$ of the beam also contributes to the beam size. This part has to be quadratically subtracted by

$$\varepsilon_x = \frac{1}{\beta_x(s)} \left[ \sigma_x^2(s) - \left( D_x(s) \frac{\Delta p}{p} \right)^2 \right]$$

where $\Delta p/p$ is the momentum spread and $\beta(s)$, $D(s)$ and $\Delta p/p$ can be determined sufficiently precisely in the case of circular accelerators by common diagnostic methods. Therefore, a measurement of the beam profile is often sufficient to determine the beam emittance. Minimally invasive instruments are needed to avoid destroying the circulating beam. These instruments are discussed in Section 4. In linear machines and transport lines the optic parameters depend on the incoming beam and the emittance has to be determined by measuring all elements of the σ-matrix or ⟨x⟩ and ⟨x'⟩. Methods and instruments for this purpose are discussed in this section. Owing to their nature of measuring the angular and spatial beam distributions at the same time, these types of instruments are fully destructive to the beam. They cannot be used in full-power high-intensity beams. They are installed in nearly all machines, however, to enable a measurement of the optic parameters at low current to tune the machine and to prove its health.

## 3.1 Screens and harps

Most typical devices for emittance measurement use fluorescent screens or SEM harps as detectors, therefore a brief introduction is given first. Note that both types are intercepting devices with cannot be used at full beam intensities. Also note that the energy deposition of heavy ions at very low energies is at a maximum and the penetration depth is very small. Therefore, the energy is deposited in a very small volume creating extreme hot spots in the material.

### 3.1.1 Screens

The observation of beam profiles on fluorescent or scintillation screens is one of the oldest and most common diagnostic techniques. Together with modern TV cameras and imaging processing this technique offers simplicity, reliability and high resolution. The resolution is limited by the grain size of the material and by optical effects, mainly due to the depth of field when viewing the screen under 45°, but the thickness of the material itself also leads to light collection in the depth which disturbs the image [48]. A thin phosphor layer (e.g. P43 or P46) with a small grain size (e.g. 5 µm) reduces this effect. Quite a number of different materials exist which are used as viewing screens in particle beams [49]. The sensitivity of the materials covers more than four orders of magnitude and decay times from nanoseconds to seconds are possible. Table 2 gives an overview of some common materials.

**Table 2:** Screen sensitivities for hadron beams, from Refs. [49–51]

| Material | Relative sensitivity | Decay time (10%) | Maximum emission |
|---|---|---|---|
| $Al_2O_3$:Cr | 1 | > 20 ms | 700 nm |
| CsI:Tl | ≈200 | 900 ns | 550 nm |
| Li Glass:Ce | ≈0.05 | 100 ns | 400 nm |
| Quartz $SiO_2$ | ≈0.005 | 1 ns to 10 ns | 600 nm |
| P43 ($Gd_2O_2S$:Tb) | ≈1.5 | 1 ms | 545 nm |
| P46 ($Y_3Al_5O_{12}$:Ce) | ≈0.2 | 300 ns | 530 nm |
| ZnS:Ag | ≈0.1 | 200 ns | 450 nm |

The major problem of viewing screens systems is the radiation damage of the components, the screen and the camera. Studies in Refs. [51, 52] showed that the light yield of some materials depends strongly on the integrated charge on the screen. In addition to the decrease of light it also results in broadening of the image due to the non-linear response of the screen. At high beam currents the temperature of the screen increases at the beam spot due to energy deposition in the material. Some dependence of the light yield on the temperature is reported[5] [51], also resulting in a non-linear response. Additional care has to be taken due to possible saturation of the light yield inside the material at high beam intensities [53, 54]. Unfortunately for the moment there is no clear recommendation for "the best" material for the use in high current applications and further studies are still needed.

For high-energy beams the use of optical transition radiation (OTR) became reasonable [55]. The thickness of OTR screens can be about one order of magnitude less than conventional screens. OTR is linear over a large range and it is very fast. Its use for bunch length measurements is limited due to the weak signal for hadrons. Some aging of the signal has been observed after $10^{19}$ protons, but the signal was still useable [56]. Many more details can be found at the Workshop of Scintillating Screen Applications in Beam Diagnostics [57].

CCD cameras suffer from radiation damage in radiation fields like most semiconductor devices. The screen itself is a source of scattering of beam particles and nuclear interactions. The resulting radiation might consist of energetic α, β, γ and neutrons which induce ionizing (electron–hole creation) and non-ionizing (e.g. displacement of atoms) processes in the semiconductors. The consequence of these processes are permanent damage of the material resulting in various effects such as increasing dark current, change of bias voltage or even complete malfunctions. CCD cameras often reply with a degradation of contrast, blind CCD pixels or gain variations. CCD cameras might become unusable already after about 10–20 Gy (see Ref. [58]). Old-fashioned vidicon cameras were somewhat more radiation resistant but they hardly exist anymore. Radiation-resistant CCD cameras are useful up to some tens of kilograys. The dynamic range of a screen + CCD station is typically limited by the dynamic range of the camera or the image processing (typically 8 bit), as long as saturation of the screen can be avoided.

*3.1.2    SEM harps*

SEM harps consist of stretched metallic wires orientated perpendicular (horizontal or vertical) to the beam. Each individual wire is connected to an electrical vacuum feedthrough and an amplifier. The vertical oriented harp measures the horizontal beam profile and vice versa. Beam particles hitting a wire create secondary electrons of 20 eV to 100 eV from its surface. The secondary electron yield *Y* is described by the Sternglass formula [59]:

$$Y = 0.002 \text{ cm MeV}^{-1} \cdot \frac{dE}{dx}$$

where *dE*/*dx* is the stopping power of the particle in the material. The secondary electron emission (SEE) efficiency ε is defined as the ratio between the number of secondary electrons and the number of traversing particles. It varies between about 300 % for low-energy protons (e.g. 40 keV) to about 2 % for minimum ionizing particles for most common metals. The SEE current in each wire is proportional to the number of beam particles hitting the wire and is linear over many orders of magnitude. An appropriate profile harp consists of defined spaced wires of defined diameter and material.[6] The spacing and the diameter of the wire defines the resolution of the instrument. Typical values of both are between 20 µm and about 2 mm. Well-suited wire materials are tungsten and titanium due to their good mechanical and thermal properties, but carbon and aluminum are also used. Using low-Z materials and thin foils instead of wires has the advantage of lower beam losses due to the interaction with the wire and less heating [60, 61]. Sometimes the wires are gold plated to improve the long-term stability of the SEE yield [62] or CsI plated to increase the SEE yield [63]. A negative bias voltage on the wires [62] or positive collection electrodes [64] avoids recollection of the secondary electrons by the same or another wire. The dynamic range of a SEM harp is limited by the electronic noise on the low beam current end and by thermal electron emission due to heating of the wire at the high beam current end. High dynamic range signal processing can be done by logarithmic amplifiers with a dynamic range of $10^7$ (see Ref. [65]) or by selectable gain amplification [66]. A parallel readout of all wires enables a profile measurement of a single passage. Special care has to be taken for high brilliance beams to avoid too much heating of the wires. In addition to thermal electron emission, the wire itself or its support (e.g. soldering) may melt and the elongation of the wires changes with heat [67]. After a large integrated number of particles had crossed the grid some reduction, up to 50 %, in the secondary emission efficiency had been observed, especially for aluminum and gold-plated materials [68].

---

[5] Note the high-energy deposition of ion beams.
[6] Note that these parameters might depend on the position within the harp, e.g. for optimum measurements of the beam core and the tail.

Only about 10% or less of the beam area is covered by the wires. Even the wires can be made of very thin strips of light material. Therefore, such a SEM harp (or a thin screen) is quite transparent for not too low-energy beams and successive harps or multiple beam passages are possible. A turn-by-turn profile measurement in a circular accelerator enables injection mismatch studies by observing beam width oscillations [69, 70].

### 3.1.3 *Emittance measurement by slit + screen/harp and pepperpot methods*

A direct measurement of the beam emittance without knowing the Twiss parameters is possible in low-energy hadron beams. Those particles which have a penetration depth of a centimetre or less can be stopped by a simple metallic plate. A small transversal (either horizontal or vertical) slit of height $h_{slit}$ in this plate selects a beamlet which shines on a screen [71] or harp [72] monitor after a drift distance $d$. The width of the measured beamlet is defined by the divergence of the beam $x'$ at the slit position $x_n$, on the distance $d$ and on the resolution of the system defined by $h_{slit}$ and the resolution of the screen/harp device. The height of the signal depends on the bunch current $I_0$ and on the current distribution in the bunch (the beam profile). The slit is scanned across the beam profile and for each position $x_n$ a beamlet distribution $I(x_n, x')$ is collected. The two radii $r_{1,2}$ of the distribution give the angular spread of the beam at the position $x_n$ by

$$x'_{1,2} = r_{1,2} / d$$

The radii of the distribution are defined by the amount of current included in the emittance ellipse. This is illustrated in Fig. 8. For each $x$ the corresponding $x'_{1,2}$ are plotted in a contour plot where the included area $A_x$ is then the emittance of the corresponding current level (see Fig. 9):

$$\varepsilon_x(x,x') = 1/\pi \int dx\, dx' = 1/\pi \cdot A_x$$

The rms emittance of the incident beam

$$\varepsilon_{rms}(x,x') = \sqrt{\langle x^2 \rangle \langle x'^2 \rangle - \langle xx' \rangle^2}$$

can be calculated from the measurements by a method described in Ref. [73]

$$\varepsilon_{rms}^2 \approx \frac{1}{N^2} \cdot \left[ \sum_{j=1}^{p} n_j \left( x_{nj} - \bar{x} \right)^2 \right] \cdot \left[ \sum_{j=1}^{p} \left( n_j \cdot \left( \frac{\sigma_j}{d} \right)^2 + n_j (\bar{x}'_j - \bar{x}')^2 \right) \right]$$

$$- \frac{1}{N^2} \left[ \sum_{j=1}^{p} n_j x_{nj} \bar{x}'_j - N \bar{x} \bar{x}' \right]^2$$

where $N$ is the total number of particles crossing the slit during a scan, $x_{nj}$ is the $j$th slit position, $p$ is the number of different slit positions, $n_j$ is the number of particles passing the though the $j$th slit position (proportional to the signal intensity), $\bar{x}$ is the mean position of all beamlets, $\bar{x}'$ is the mean divergence of all beamlets, $\bar{x}'_j$ is the mean divergence of the $j$th beamlet and $\sigma_j$ is the rms divergence of the $j$th beamlet. The same formalism is true for the $y$-plane. In this case the slit has a perpendicular orientation.

The procedures described above neglect the finite resolution of the device. A small $h_{slit}$ and small step sizes of the slit positioning (even when using more than one slit) are necessary to improve the resolution. When using a grid as detector its resolution can be improved by scanning the grid as well. A position of the device in a region with high $x'$ is most helpful for a good resolution. In high-density beams care has to be taken that space charge in a beamlet might extend the angular spread of it on the way to the detector. Some recent studies of the influence of the slit geometry can be found in Ref. [74].

The Allison Scanner uses a FC instead of a position sensitive detector behind the slit, while the angle distribution $x'$ is measured by an electric sweep. It is limited to small energy ion beams (<100 keV) due to the limit in sweeping voltage.

The scanning of the whole beam size in both planes needs quite a lot of time with very stable beam conditions. To overcome the scanning procedure a plate with thin holes in defined distances (pepperpot) produces a lot of beamlets in the $x$, $y$ plane (see Fig. 10 and Ref. [75]). Therefore, the whole phase space can be measured with one single shot onto the pepperpot mask (see Fig. 10). This will reduce the heat load of the plate drastically with respect to slowly scanning slits.

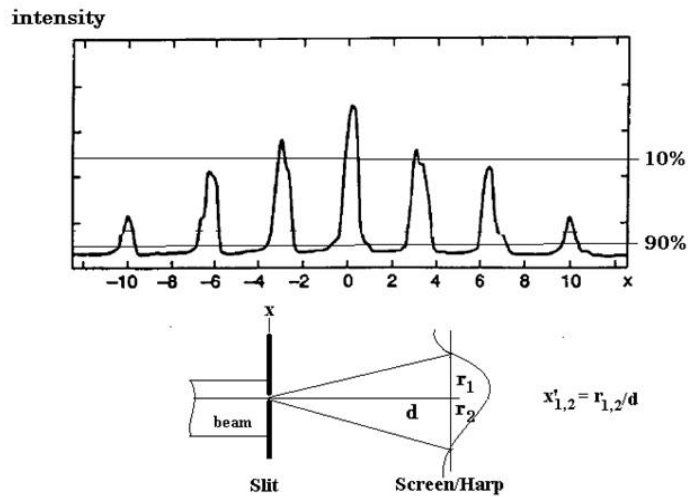

**Fig. 8:** Beamlet distributions at seven different $x$ positions of the slit. The horizontal lines indicate approximately the 10% and 90% level of the beam current taken into account for the emittance calculation. Here $x'_{1,2}$ is then defined by the corresponding radius $r_{1,2}/d$.

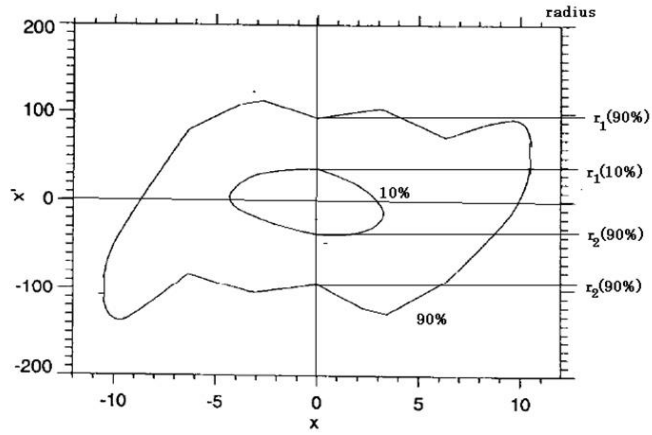

**Fig. 9:** Contour plot of the *x, x'* phase space. The measured radii at a slit position of *x* = 0 are indicated on the right-hand side.

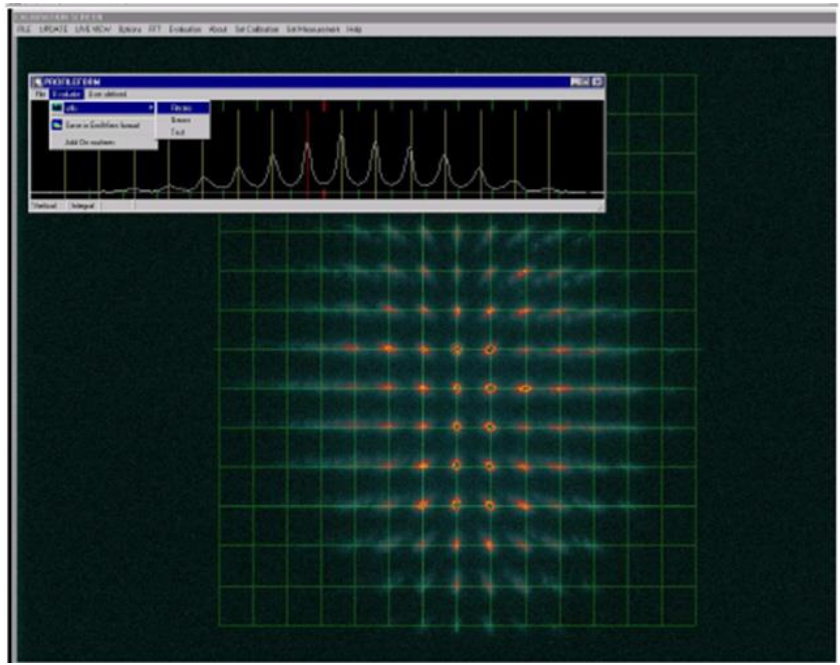

**Fig. 10:** Screen shot from the pepper-pot device for an Ar beam and, as an insert, the projection onto the horizontal plane (reproduced from Ref. [75])

### 3.2 Emittance measurements by quadrupole variation or three screens/harps methods

In a beam transport system the sigma matrix σ is transformed from one point $s_0$ to another $s_1$ by

$$\sigma(s_1) = M \sigma(s_0) M^{\mathrm{T}}$$

where *M* and $M^{\mathrm{T}}$ are the transport matrix between the two points and its transpose, respectively. In a dispersion-free beam transport all are 2 × 2 matrices and the measured width $\sigma_{measured}$ at ($s_1$) is then

$$\sigma_{measured}^2(s_1) = \sigma_{11}(s_1) = M_{11}^2 \sigma_{11}(s_0) + 2 \cdot M_{11} M_{22} \sigma_{12}(s_0) + M_{12}^2 \sigma_{22}(s_0)$$

The matrix elements $M_{ij}$ are known by the elements of the transport system between the two points. The three unknown elements of the $\sigma_{ij}(s_0)$ matrix can now be resolved by at least three different measurements, either by three profiles at three different locations in the transport system or by three profiles with three different transport optic settings, e.g. by variation of the focusing strength of a quadrupole. The matrix elements $M_{ij}$ of different measurements have to differ significantly to solve the linear system, therefore the different locations have to have enough phase advances or the focal strength of the quadrupole has to be varied over a large enough range.

Typically the beam profiles are measured by thin screens or harps. This has the advantage of small and almost negligible beam blow-up due to the measurement itself so that the beam can transverse a few screens. This enables a "one shot" measurement of even a single bunch while the scanning methods need a stable beam over the scanning time.

Three measurements give a unique solution but no error estimation. Therefore, more measurements, either by more stations or by more quadrupole settings are recommended. In particular, the variation of the quadrupole settings allows a quadratic fit of the square of the measured beam size versus the quadrupole gradient together with an estimate of the errors in the measurement [76].

The standard methods descript above are valid under the assumption that:

- the dispersion along the section is zero;
- the transfer matrices are known;
- no coupling is present between the two planes; and
- no space charge or other non-linear forces are present.

In particular, a dipole in the section between the monitors creates dispersion which has to be taken into account. In this case or with initial dispersion the particle trajectory vector then includes the momentum spread $\vec{x} = (x, x', \Delta p/p)$ and the $\sigma$ matrix and the transport matrix $M$ becomes a $3 \times 3$ matrix. In this case at least six measurements are necessary to determine all $\sigma$ matrix elements to resolve the emittance. The general case is discussed in more detail in Ref. [77].

The influence of space charge in high-intensity beams can be measured by observing the emittance evaluation in dependence on the intensity. As discussed before, care has to be taken not to saturate or even destroy the monitor with the beam. A detailed discussion of the equations of motion and of the emittance in the presence of space charge can be found in Ref. [78].

A full reconstruction of the four-dimensional phase space $(x, y, x', y')$ by tomographic image reconstruction of the spatial beam projections (profiles) has become a useful diagnostic tool over the last few years [79]. The quadrupole scanning technique or a set of screens along the transport line with sufficient phase advance delivers the input for the computerized tomography.

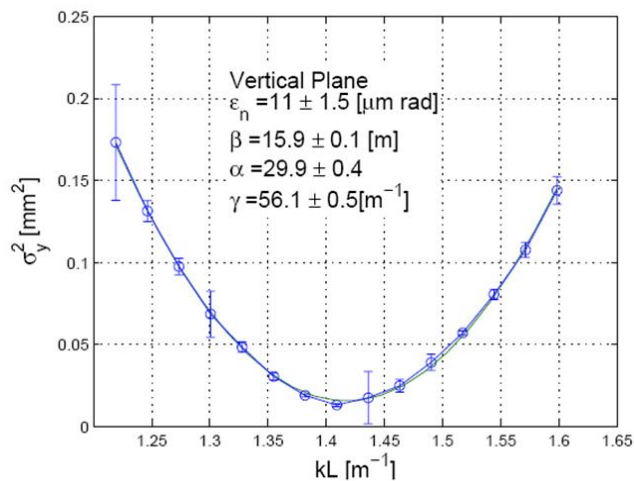

**Fig. 11:** Scan of the quadrupole gradient versus the measured beam width and fit to data (reproduced from Ref. [76])

## 3.3 Emittance measurement by tomography

Tomography is the technique of reconstructing an image from its projections; see Fig. 12. It is widely used in the medical community to observe the interior of the human body by processing multiple X-ray images taken at different angles. Beam phase space tomography reconstructs the phase space density distribution by means of one-dimensional profiles (projections) from beam profile monitors by means of a mathematical algorithm.

The main reconstruction algorithms used are [80]:

- convolution and back projection methods (FBP);
- maximum entropy (MENT) algorithm;
- maximum likelihood expectation maximization (MLEM);
- algebraic reconstruction techniques.

The "filtered back projection" or "convolution" reconstruction process is widely used because the mathematics is simple and easily programmable. For a small number of projections, however, streaking artefacts dominate the reconstructed image. The optimum algorithm depends strongly on the problem being solved. Some algorithms are better at reconstructing Gaussian distributions, whilst others are suited to detailed distributions

Some questions arise regarding the limitations of tomography technique for space charge dominated beams. The use of linear space charge forces led to inconsistent results [81].

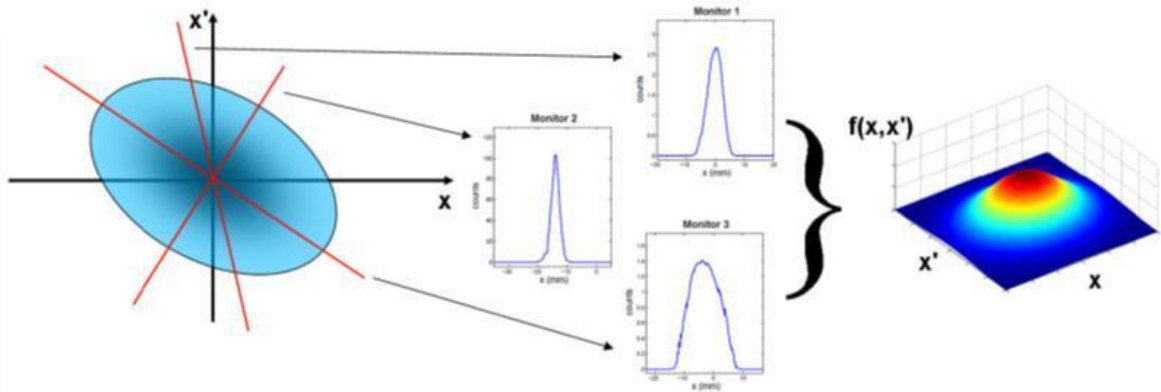

**Fig. 12:** The plot on the left-hand side represents the unknown beam transverse phase space distribution at the reference position $z = z_0$. Beam profile monitors acquire projections of the phase space onto the $x$ coordinate at different locations with a certain phase advance (middle plots). These projections are related to the beam distribution at $z = z_0$ through linear transport matrices accounting for drift space and/or quadrupole magnets. In beam tomography the profile data are employed by a mathematical algorithm to reconstruct the two-dimensional beam density distribution (right plot). (Reproduced from [82].)

## 4 Instruments for beam profile measurements

The aim of transversal beam profile measurement is to determine the transverse shape of the beam down to about $3\sigma$ to $4\sigma$. Further outside, halo measurement starts (see Section 6). Therefore, a dynamic range of $10^3$ to $10^4$ is sufficient for a single measurement. Additional constraints come from the requirement to measure the profile at different beam currents. Therefore, the profile monitor needs additional pre-gain settings to adapt to the current issue. The required spatial resolution depends on the beam size. Typical beam dimensions of hadron beams are in the millimetre range so that a resolution of 100 µm is sufficient in most cases.

The main motivation for beam profile measurements is to understand the beam dynamics in the machine and, in conjunction with that, to minimize beam losses along the accelerator (see also Section 5). There are many sources which can drive a blow-up of the beam core such as space charge, scattering, mismatch, resonances, etc., which can be observed by profile monitors. In a chain of successive accelerators profile monitors are indispensable to measure the (normalized) emittance evolvement at each step of the chain.

In contrast to the destructive emittance measurement (Section 3) the profile measurement needs to be minimally invasive for two reasons: (1) to avoid influencing the beam and (2) to avoid destroying the monitor.

### 4.1 Wire scanner

Wire scanners are used in many accelerators as a standard device for beam profile measurements. The device sweeps a thin wire through the beam while plotting a signal which is proportional with the number of particles interacting with the wire versus the measured position of the wire (see Fig. 13). Optical rulers can determine the position of the wire with a resolution of 1 μm to 2 μm, but only at a speed of ≤1 m s$^{-1}$. Higher speeds (e.g. 5 m s$^{-1}$ [83] and up to 20 m s$^{-1}$ [84, 85]) are required for intense and high brilliant beams in circular machines for two main reasons:

(1) Reducing the heat load of the wire due to the interaction with the beam; the heat load is inversely proportional to the speed [86].
(2) Reducing the emittance blow-up of the beam due to the wire interaction since the emittance blow-up is also inversely proportional to the speed of the wire [87].

High speed is realized by circular movement of the wire which reduces the position resolution and therefore the profile resolution to 10 μm to 100 μm. The speed of a linear wire scanner is mainly limited by the vacuum bellow stress property which limits the acceleration of the mechanical feedthrough to a few *g*. New methods for fast scanners with high resolution are under study [88].

Light materials with long radiation length are preferred to reduce the emittance blow-up and to minimize the energy deposition in the wire. On the other hand a high melting point is preferred to extend the lifetime of the wire. For that reason a thin (7 μm to 20 μm) carbon wire is often a good choice due to its high melting temperature of about 3500 °C and its excellent mechanical stability.

The main cooling processes are thermionic emission and black-body radiation. Both become important at temperatures well above 3000 °C (see Ref. [89]). This is true for the high duty cycle interaction in storage rings. At low duty cycles the heat transmission along the wire becomes dominant [90]. The calculation of the heating of the wire must include the effect of the emission of secondary particles such as delta rays and others. This reduces the amount of deposit energy in the wire by up to 70 % [86, 90], depending on the beam energy. Sublimation of the wire material takes place even before the melting temperature is reached, however, and reduces the material at each scan [89]. The heating of the wire often limits the use of wire scanners in high-intensity and high-brilliance beams to low currents only.

In linacs with low duty cycle a fast scan does not make sense since the bunch train (pulse) might be too short to allow a scan within one pulse.[7] Therefore, the wire has to crawl though the beam and the profile is acquired pulse by pulse. A few data points per 1σ beam width should be the minimum to obtain a useful profile. To avoid instantaneous overheating of the wire the charge of each pulse has to be limited, to avoid an integral overheating the repetition rate of the pulses has to be limited [90, 91]. In particular, low-energy ions will deposit huge amounts of energy even in thin wires, so that their use is very limited in such accelerators. The use of wire scanners for partially stripped ions is excluded since their interaction with the solid wire will change the charge state of the ions.

---
[7] Exception: Superconducting FEL Linacs might allow bunch trains of some hundred μs but with beam size of less than 100 μm

The signal from the beam–wire interaction can be detected with two different methods:

(3) Detection of scattered beam particles[8] outside the vacuum chamber. At energies above the pion threshold (>150 MeV) the beam particles mainly interact with the wire by multiple scattering and nuclear interactions. Beam and secondary particles with large scattering angles will hit the vacuum chamber and create a shower which is detected by fast loss monitors, e.g. scintillation counters. Monte Carlo studies are most helpful to find an efficient position for the detector somewhere downstream of the scanner. Note that the signal can depend on the wire position, especially when using asymmetric detector positions at large beam sizes [92]. A fast scintillation counter is able to resolve single bunches in a train or in a stored beam. While in a linac the beam profile is a composition of many (desirably similar) bunches, the profile of each individual bunch can be measured in a storage ring [93].

(4) Secondary electron emission current created by beam particles entering and leaving the conducting wire (see also Section 3). This method is often used in low-energy beams where the scattered particles cannot penetrate the vacuum pipe wall. In this low-energy regime the stopping power of the wire forces the hadron beam particle to stop in the wire, so that the signal is a composition of the stopped charge (in the case of $H^-$: proton and electrons) and the secondary emission coefficient. Therefore, the polarity of the signal may even change, depending on the beam energy and particle type [94, 95]. When using multiple wires on one scanner, too narrow wires may cross-talk by receiving the electrons from the other wire. If the temperature of the wire exceeds the thermionic threshold the emission of thermal electrons starts to superimpose the secondary electron emission signal. Therefore, the useable temperature range is limited by that threshold for the secondary electron emission method.

Since the signal generation during a scan is a sampling process, the beam should be quite stable during the scan. In case of linacs, the beam current and position have to be correlated with the signal for each bunch.

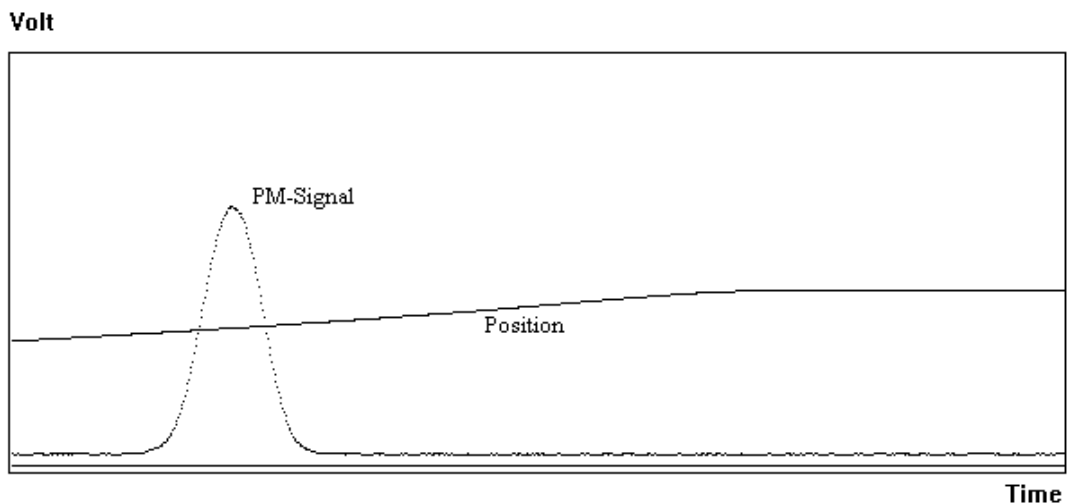

**Fig. 13:** Signal from a scintillation counter (PM-signal) and the measured position of the wire by a potentiometer versus the scan time. The real beam profile is a result of plotting the signal versus the position.

---

[8] or Bremsstrahlung in case of an electron beam

## 4.2 H⁻ laser scanner

The use of photodetachment of a H⁻ beam by laser photons was first used in Ref. [96] but to measure longitudinal H⁻ beam parameters. In Ref. [97] a well-focused laser beam was proposed to scan the intense H⁻ beam at the LAMPF accelerator. This nearly non-invasive method has the advantage neither to produce emittance blow-up nor intrinsic wire heating but it is applicable only for H⁻ beams. The cross-section for photodetachment of H⁻ ions is large enough (some $10^{-17}$ cm$^2$; see Ref. [98]) to neutralize a fraction of the beam-slice crossed by the laser. The number of photodetached electrons (and neutral H⁰) is proportional to the beam density and the laser energy density. The cross-section has a maximum for photon wavelength around λ = 1000 nm (= 1.2 eV) so that the second electron (binding energy 13.6 eV) will not be stripped by those laser photons. The 1064 nm light from a Nd:YAG laser is very close to the optimum wavelength, but for relativistic H⁻ particles the Lorenz boost has to be taken into account which increases the photon energy in the rest frame of the H⁻ ($E_{CM}$) by

$$E_{CM} = \gamma \, E_{YAG} \, (1 - \beta \cos \Theta)$$

where Θ is the crossing angle between the laser and beam. For a H⁻ beam with $E_{kin}$ = 1 GeV this reduces the cross-section to about 70 %, but the photon flux also receives a Lorenz boost in the same way keeping the photodetachment yield nearly constant $0.2 \leq E_{kin} \leq 1$ GeV. A detailed calculation of the photodetachment yield is discussed in Ref. [99].

A $Q$-switched Nd:YAG laser (up to few hundred millijoules) can be synchronized with the ion bunches. Since the laser pulse is typically much longer (some nanoseconds) than the bunches, an injection seeder is required to smooth the temporal laser pulse profile [100]. The bunch position, bunch current, laser shot-to-shot variations and drifts have to be monitored during the scan and normalized to the results. The laser focus has to be significantly smaller than the H⁻ beam size and its Raleigh length correspondingly larger to ensure a clean measurement. The laser is scanned across the beam by a motorized mirror system. Since a laser beam can be transported over long distances one laser can serve many scanning stations, e.g. the 9 stations along about 300 m at the SNS Linac are served by one laser [101].

Both, the liberated electrons and the remaining H⁰ can be detected to measure the beam profile (see Fig. 14):
(1) The neutral H⁰ reduce the bunch current. This is measured by a FCT while its amplitude is plotted versus the laser beam position.
(2) The liberated electrons are bent by a small dipole field into a FC.

Since the electron energy is only a few kiloelectronvolts a dipole field of 50–150 G is sufficient and its feedback on the H⁻ beam is quite small. To collect the electrons diffused by space charge a wide area FC is required, located downstream near the laser interaction point. A biased collector (≈200 V) with a repeller grid in front ensures suppression of background and secondary electron emission. A second electrode in front of the interaction point can be helpful to collect (background) electrons created by Lorentz or residual gas stripping. Beam losses near the monitor are the remaining source of background which should be avoided to get a high dynamic range of the measurement. The repeller grid is also used to measure the energy distribution of the electrons, which is a sum of their initial energy and the energy gained by the space charge of the beam [102].

The direct use of H⁰ enables a direct emittance measurement using the laser as a "slit". After bending the H⁻ beam the neutral H⁰ remain on a straight line where a screen or grid measures their distribution [103]. The laser "slit" has the advantage of avoiding the thermal problems that exist in conventional slit-grid monitors (see Section 3). To reduce the background of H⁰ produced in front of the laser, Ref. [104] has the interaction with the laser in the middle of a dipole so that the laser neutralization takes place after a small bend and only those H⁰ are collected on the target.

A laser pulse much shorter than the temporal current distribution of the bunch enables also a bunch length measurement. A mode locked Ti-Sapphire laser reaches a wavelength of 950 nm and pulse lengths from picoseconds down to tens of femtoseconds. For the typical some ns long H-bunches the timing requirements are somewhat relaxed. The transversal size of the laser spot and of the beam should be roughly the same and stable in their positions. The laser pulse is locked to the RF frequency and its phase is scanned across the bunch length. Measurements at SNS are reported in Ref. [105].

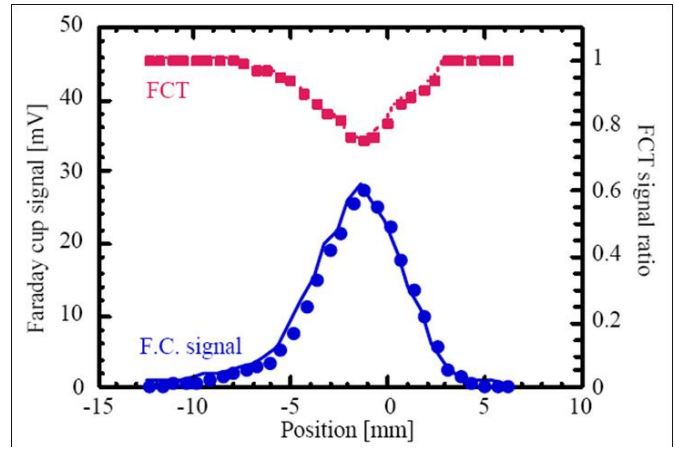

**Fig. 14:** FC signal and FCT signal from a laser scan (reproduced from Ref. [102])

### 4.3 Ionization profile monitor

Residual gas atoms or molecules are always present in the vacuum system of every accelerator. They fill the beam pipe with a homogeneous distribution, typically with a pressure of $10^{-6}$ mb to $10^{-9}$ mb. Most of the residual gas components are $H_2$ molecules. Assuming a mean energy of about $E_{ion} = 90$ eV needed to ionize a molecule of the residual gas (95 eV for $H_2$ only), the amount $N$ of ionized particles can be derived from the Beth–Bloch Formula ($dE/dx$) valid for the individual pressure:

$$N = \frac{1}{E_{ion}} \cdot \frac{dE}{dx} \quad [\text{cm}^{-1} \text{ particle}^{-1}]$$

More accurate measurements of the ionization cross-sections were performed in Ref. [106]. The simple model from the formula above results in $N \approx 40$ ion–electron pairs per centimetre at a $H_2$ pressure of $10^{-9}$ mb and a passage of $10^{13}$ minimum ionizing particles.

The IPM separates the resulting electron ion pairs by the use of an extraction field $E_{ext}$ perpendicular to the beam axis. Typical values for $E_{ext}$ lie between 50 V mm$^{-1}$ and 300 V mm$^{-1}$, depending on the gap between the electrodes, practical power supply and space charge distortion (see below) considerations.

Field-forming electrodes and their careful design guarantee a highly parallel field so that the electrons or ions are projected onto the readout plane (see Fig. 15). Extended electrodes [107] and/or coated walls are useful for cleaning the environment from secondary electrons and ions not generated in the extraction volume. Most existing IPMs are now using one or two micro-channel plates (MCPs) inside the beam vacuum to amplify the resulting current. Just after the MCP the amplified current is collected by a phosphor screen or multi-anode strips. Early examples of these system can be found in Ref. [107] (phosphor screen) and Ref. [108] (strips); the first IPM was described in Ref. [109] but without using a MCP. An internal amplification can also be achieved by using a gas curtain [110] or a gas bump [111] in the monitor. The gain of a single MCP reaches up to $10^3$ while a double stack

(chevron) reaches up to $10^6$. Since the distance and diameter of the microchannels are of the order of 10 μm, a MCP does not distort the projected beam profile significantly as long as the width is larger than some 100 μm. The secondary electrons at the output of the MCP are accelerated by a second electrical field onto a phosphor screen or a position-sensitive anode configuration.

The phosphor screen is viewed by a standard CCD video camera which provides sufficient resolution and sensitivity in most cases [112]; see Fig. 16. A carefully designed optic is important, however, to achieve a good resolution [113]. The video frame rate limits the bandwidth of this readout to 50 (60) Hz which is sufficient for storage rings, but not for cycling synchrotrons or linacs. A fast readout can be achieved by position-sensitive (silicon) photomultipliers or photodiode arrays [114, 115]. The use of a fast decaying phosphor type (e.g. P47) is then required.

A very fast readout can be achieved by using anode strips as a collector of the MCP electrons. The separation of the strips defines the spatial resolution. A pitch of down to 250 μm is possible providing sufficient resolution. Connecting each strip via a charge-sensitive amplifier to a fast ADC enables turn-by-turn [116] and even bunch-by-bunch [117] resolution. The use of a resistive plate or wedge-and-stripe anodes [118, 119] can reduce the number of channels and vacuum feedthroughs. Since this is based on the detection of single particles, it disables the very fast readout opportunity. On the other hand it can improve the resolution up to the limit of the MCP by applying high statistics.

IPMs in high-intensity accelerators suffer from the high space charge of the (bunched) beam, exceeding the extraction field $E_{ext}$. The space charge disturbs the exact projection of the beam profile by bending the trajectories of the ions and electrons [116]. In particular, the light electrons get such a large kick that a profile measurement becomes impossible. Applying a magnetic field parallel to the extraction field forces the electrons to spiral around the magnetic lines with the cyclotron radius $r_c = m_e v_\perp / eB$. The radius depends only on the initial transverse kinetic energy defined by the kinematics of the ionization process and is below 50 eV for 90 % of the electrons. A magnetic field of about 0.1 T is then sufficient to keep the radius (and therefore the monitor resolution) below about 0.1 mm (see Ref. [120]). Different design with permanent magnets [121] and electromagnets have been realized [117], but note that some electrons may reach much higher kinetic energies which lead to tails of the distribution produced by intrinsic effects and not by the beam halo. The extraction field $E_{ext}$ and the magnetic field $B$ have to be compensated by opposite fields close to the monitor to minimize any influence on the beam.

By changing the polarity of the extraction field $E_{ext}$, it is also possible to collect the ions on the MCP. The heavy ions are not so strongly affected by the space charge and the distortion can be analysed and subtracted. A precise correction includes also terms from the collision impact and from the thermal movement of the residual gas molecules. The collision impact on the ions is quite small but the thermal velocity spread contributes already with a profile broadening of approximately 200 μm to 300 μm, depending on $E_{ext}$ and the monitor geometry. The broadening of the measured profile due to space charge is reasonable at lower bunch densities but can exceed by far the real beam width at high bunch density. Under those conditions a correction might become useless. Since the broadening depends on all bunch parameters, detailed calculations and simulations are required to estimate the error contributions in detail [122–125].

A well-known problem of MCPs, phosphor screens and anodes is their aging with the amplified charge. This leads to a reduced gain just in the centre of the beam distribution and broadens the measured profile. Therefore, a continuous monitoring of the gain distribution is required. Various methods are in use or discussed, such as an α source [118], an electron generator plate [126], a tungsten filament emitter [127], an ultraviolet lamp [128] or a motorized 90° flip of the MCP [129]. A simple way to prove the aging is to steer the beam to an unused part of the sensitive area.

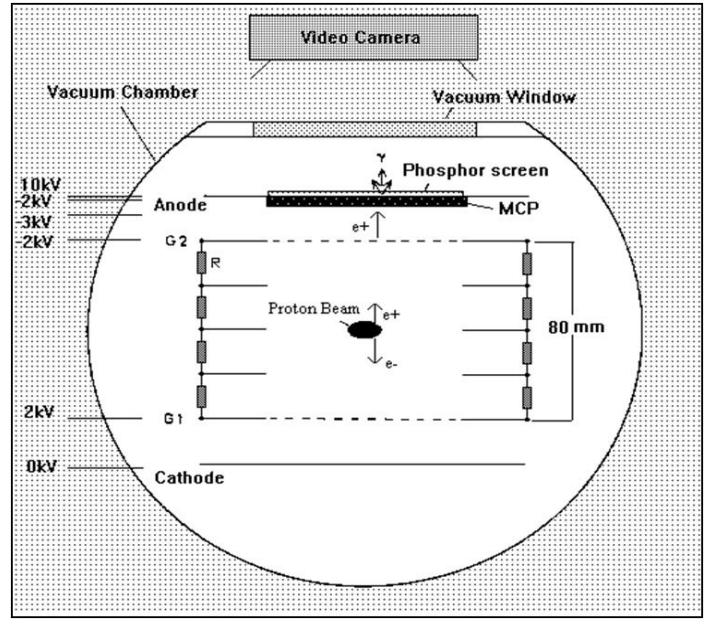

**Fig. 15:** Sketch of an IPM (in the ($y$, $z$)-plane) with MCP and phosphor screen. The extraction field $E_{ext}$ is applied between G1 and G2 (grids). A resistive network R and the field shaping electrodes provide a homogeneous field distribution. The voltage configuration shown is an example to collect residual gas ions on the MCP.

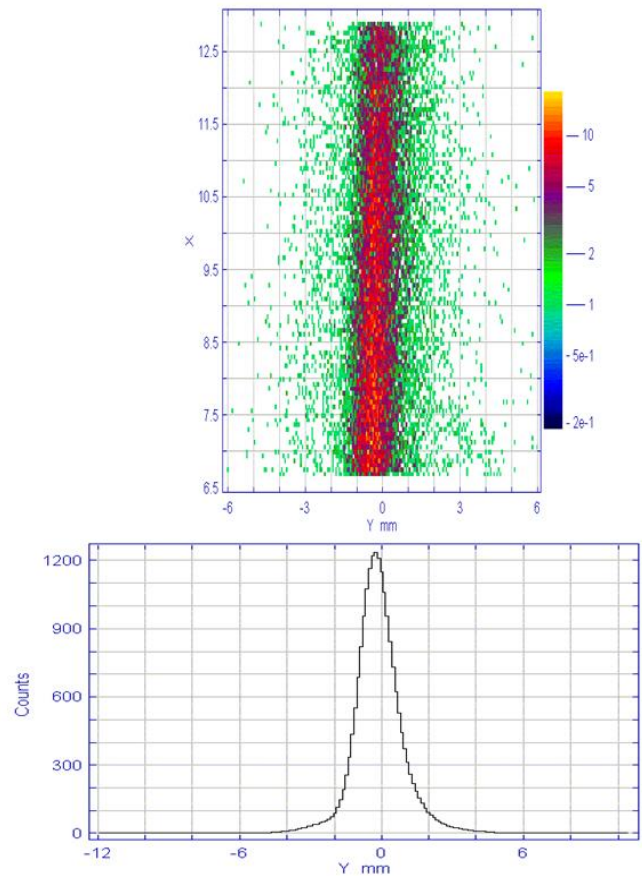

**Fig. 16:** Image of the beam on the phosphor screen (upper part) and its projection on the $y$-plane of the beam (lower part) (reproduced from [118])

### 4.4 Gas scintillation

The molecules of the residual gas are not only ionized, but are also excited by the beam particles. During their de-excitation they emit light in the visible range (depending on the gas type). Assuming a homogeneous distribution of the gas atoms in the vacuum, the focused light distribution represents the beam profile. The cross-section for gas scintillation is much smaller than for gas ionization; only a small percentage of the ionization loss is converted into detectable visible light [130]. This still does not include the fact that the light is emitted into the full solid angle while the detector only covers a fraction of it. Also not included is the finite sensitivity of a photocathode of a position-sensitive detector. Owing to this small efficiency typically a pressure bump is needed to increase the signal to a sufficient level. Nitrogen ($N_2$) is often used for local gas bumps since the vacuum system can efficiently pump this gas. The dependence of the cross-section on the ionization loss (Bethe–Bloch formula) has been proven over a wide energy range with proton and ion beams, as well as its linear dependence on the $N_2$ gas pressure [131, 132], but note that a partially stripped ion beam cannot handle significant added gas pressure.

If the excited residual gas atoms are still neutral atoms, they not affected by the space charge of the beam and this method is quite suitable for high-intensity beams and bunches, but the excitation cross-section is highest for the ionized state $N_2^+$ (see Ref. [133]) with a half lifetime of the de-excitation of about 60 ns (see Refs. [131, 134]). The movement of the excited ions during this time depends mainly on the space charge of the bunch. Simulations for LHC have shown drifts of some hundred micrometres, creating long tails in the measured beam profile [135]. Using Xe as a working gas with a lifetime of 6 ns ($Xe^+$) can improve this situation [136] but a much lower cross-section has to be taken into account. Other contributions which further broaden the measured profile of the order of some tens of micrometres include [135, 137]:

- thermal movement of the ions;
- momentum exchange during ionization;
- finite impact parameter;
- secondary electrons create additional excitations far away from the beam.

There is some experience with other gases used for a gas scintillation monitor. Xenon and other noble gases were studied in Refs. [131,136, 138], with a first result that He is excluded due to large tails in the beam profile.

High-sensitivity light detectors with position resolution are necessary for this type of monitor, even with a pressure bump. High gain image intensifiers, intensified CCD cameras or position-sensitive photomultiplier tubes (PMTs) are typically used. With practical gas bumps signal integration over many turns is still required to get a useful signal so that turn-by-turn or single shot profile measurements are excluded. All devices need good protection against radiation from adjacent beam losses to reduce background signals as well as radiation damage of the detector.

## 5 Beam loss measurements

A serious problem for high-current and high-brilliance accelerators is the high power density of the beam. A misaligned beam is able to destroy the beam pipe or collimators and may break the vacuum. This fact makes the BLM system one of the primary diagnostic tools for beam tuning and equipment protection in these machines. In addition to the task of machine protection the BLM system has more major goals:

- It should limit the losses to a level which ensures hands-on-maintenance of accelerator components during shutdown and it should limit the radiation outside the accelerator shielding. The hands-on limit has been found approximately between 0.1 W m$^{-1}$ and 1 W m$^{-1}$ (see Refs. [139, 140]). A value of 1 W m$^{-1}$ corresponds to 1 GeV·nA m$^{-1}$; note that the limit of losses shrinks with beam energy.

- Ground water activation and radiation damage to components may put additional constraints on tolerable beam losses [139].
- Detecting the physical locations of a beam loss within a certain resolution in space. Often the resolution is limited by the spacing of the individual BLMs.
- Determination of the fraction of the lost particles relative to the beam, within a certain time interval.
- The system should be sensitive enough to enable machine fine tuning and machine studies with the help of BLM signals; sometimes even at low beam intensity to avoid high losses and/or during machine commissioning and at various energies during acceleration. This includes the comparison of the detected loss with computer models (Monte Carlo and beam tracking programs) and the analysis of the behaviour.

Therefore, one of the main issues of a BLM system is its very high dynamic range. It has to deal with two different types of losses; the regular losses which are unavoidable but suitable for beam diagnostics and the uncontrolled losses which generates additional radiation and risks [141].

Uncontrolled losses may occur with a fast transient, therefore the reaction time of the BLM system has to be matched to the transient time. In linacs even a bunch-by-bunch loss measurement is required while in (superconducting) storage rings about 0.1 ms to 1 ms are sufficient [142–144]. An integration of the signal over the required period is compared with a predefined threshold to generate alarm signals in case it exceeds the threshold. The threshold of tolerable beam losses depends on the specific requirements of the adjacent accelerator, e.g. quench limits, heating, radiation, residual activation, background, etc. Dangerous conditions are defined by the acceptable energy deposition of the lost particles and its adjacent shower in sensitive materials of the accelerator environment. Monte Carlo simulations are most helpful in calculate the thresholds for each specific BLM location as well as to calibrate the response of the BLM in terms of lost particles [145, 146].

Regular losses might occur continuously during operational running and correspond to the lifetime/transport efficiency of the beam in the accelerator. The lowest possible loss rate is defined by the theoretical beam lifetime limitation due to various effects, like residual gas scattering, diffusion, space charge, etc.; controlled losses due to scraping, beam extraction and injection (stripping foil), collision, etc. also fall into this category. These losses should be localized on the collimator system or on other known and properly designed aperture limits. At these locations, the measurement of losses can also be used for machine diagnostic purposes (in addition to their protection task), e.g. for optimizations of injection, lifetime, beam transport, background conditions and residual activation as well as for tail and tune scans, for measurement of diffusion processes and much more. For details see Refs. [141, 147].

BLMs should be localized in areas with higher probability of beam losses, e.g. collimators, high dispersion regions, high-β amplitudes. Different types of BLMs are used sometimes at the same location to extend the dynamic range of the system: sensitive BLMs to measure small losses and more insensitive ones to cover the high loss rates [142, 148]; or to cover different time scales: scintillator-based BLMs for nanosecond response times and ionization chambers for microsecond response times. In particular, at beam energies below the pion threshold (< 150 MeV) the (additional) use of neutron-sensitive BLMs is useful since the charged particles hardly escape the vacuum chamber [149].

Many factors are important for the design of a proper BLM system. In particular, for high-intensity beams a common aspect is the required large dynamic range, but also the radiation resistance, saturation characteristics and more. A summary of important considerations when selecting a BLM design are listed in Ref. [150]. A detailed discussion of various types of BLMs can be found in Ref. [147]. The following discussion about BLMs will concentrate on the aspect of their dynamic range.

## 5.1 Ionization chambers

Short ionization chambers are used as BLMs in many accelerators [148, 151–154]. An ionization chamber in its simplest form consists of two parallel metallic electrodes (anode and cathode) separated by a gap of width $D$ and an applied bias voltage of some hundreds of Volts. The gap is filled with gas (air, argon, xenon[9]) of density $\rho$. The gas-filled volume between the electrodes defines the sensitive volume of the chamber. Ionizing radiation creates electron–ion pairs in this sensitive volume. The electrons can escape an immediate recombination if the electric field between the electrodes is larger than the Coulomb field in the vicinity of the parent ion. If all charges are collected the signal does not depend on the applied voltage (ionization region). The flatness of the plateau of the ionization region depends on the collection efficiency of the electrons or ions on the electrodes. In particular, at high radiation levels electrons on their way to the anode may be captured by positive ions produced close to their trajectory (by other incoming particles) and do not contribute to the charge collection. Therefore, a high voltage and a small gap $D$ are preferred to achieve a high dynamic range as well as to achieve a faster response time of the ionization chamber [152]. Electron collection times of less than 1 μs are achieved, even in large chambers, by an appropriate arrangement of the electrodes [151].

The dynamic range of an ionization chamber is defined by its upper and lower current signal. The upper limit is given by the non-linearity due to the recombination rate at high dose; the typical chamber current in such a case is a few hundred microamperes. The lower limit is given by the dark current between the two electrodes. A very careful design of the chamber is necessary to very low dark currents in the order of few picoamperes. This gives a dynamic range of up to $10^8$. Such a high dynamic range needs some special signal processing. Solutions such as variable gain amplifiers [155], logarithmic amplifiers [156], high ADC resolution [153] and current-to-frequency conversion [157] are applied.

Ionization chambers can be built from radiation-resistant materials such as ceramic, glass and metal with no radiation and time aging. Special care has to be taken for the feedthroughs and to the preamplifiers. Up to more than $10^8$ rad can be tolerated by a careful design focused on radiation hardness. Air-filled ionization chambers require virtually no maintenance, leakage in $N_2$ filled chambers is not critical, but sealed Ar-filled chambers also give very few reasons for maintenance.

An enhanced sensitivity is provided by using the internal gas amplification of an ionization chamber in the proportional regime. In Ref. [158] an internal gas amplification of $6 \times 10^4$ at 2 kV, a dynamic range of $10^3$ and a fast rise time of 100 ns were reported.

A "short" ionization chamber covers only a small part of an accelerator; therefore, a large number need to be installed to detect all losses. To overcome this problem a long, gas-filled coaxial cable has been used as an ionization chamber. Position sensitivity is achieved by reading out at one end the time delay between the direct pulse and the reflected pulse from the other end. The time resolution is about 50 ns (~15 m) for 6 km long cables, for shorter chambers about 5 ns (~1.5 m) was achieved [159]. This principle of longitudinal resolution works for one-shot (turn) accelerators (and transport lines) with a bunch train much shorter than the length of the cable. For particles travelling significantly slower than the signal in the cable ($\approx 0.92c$) the resolution of multiple hits in the cable becomes difficult. In this case and for circular and multibunch machines it is necessary to split the cable. Each segment has to be read out separately, with spatial resolution equal to the length of the unit. This is done in linacs [160–162] and in some rings and transport lines [163–165]. Since the chamber geometry is not optimized for high dynamic range, their linear range is limited to about $10^3$ to $10^4$ depending on the gas contents. Long ionization chambers made of commercial cables are simple to use, cheap and they have a uniform sensitivity. The isolation is not very radiation resistant, nevertheless these cables were used in SLAC for more than 20 years without serious problems.

---

[9] Electronegative gases ($O_2$, $H_2O$, $CO_2$, $SF_6$, etc.) capture electrons before reaching the electrode. Noble gases have negative electron affinities (Ar, He, Ne) which reduces recombination.

## 5.2 PIN diodes

One can treat a PIN diode with its intrinsic depletion layer as a "solid-state ionization chamber". The required energy to create an electron–hole pair in a semiconductor is about 10 times smaller than creating an electron–ion pair in gas. Also the density of a semiconductor is about three orders of magnitude larger than for gas (at 1 atm). Therefore, the signal created by radiation is much higher per unit path length than in a gas volume, but the active volume is much smaller: a sensitive area of 1 cm$^2$ and a depletion layer of $d$ = 100 µm to 200 µm is already one of the largest PIN diodes which are commercially available. At about $U$ = 30 V to 40 V the width $d$ reaches its maximum. The transit time and the rise time of the signal of the order of a few nanoseconds due to the small gap $d$, a high electric field $E = U/d$ and a capacitance $C$ = 10 pF to 100 pF. A dark current of a few nanoamperes is typical (due to the finite resistance between the two electrodes (p$^+$ and n$^+$) of the diode) and limits its dynamic range in this "photocurrent" mode. Modest radiation damage at $10^6$ rad [166] leads to an increase of the dark current while most of the other parameters remain unchanged. Note that not too strong magnetic fields do not influence the charge collection in PIN diodes. Therefore, they can work as radiation monitors in stray fields of magnets, e.g. in high-energy experiments [167, 168].

PIN diodes are not very sensitive to γ-radiation but highly efficient to charged particles due to their thin active volumes $P$. The (hadronic) shower created by beam losses includes a large number of charged particles. The HERA BLM system consists of two PIN photodiodes mounted close together (face to face) and readout in pulsed mode and in a coincidence circuit [169]. Thus, charged particles crossing through the diodes give a coincidence signal, while γ radiation which interacts in only one diode (already with small efficiency) does not [170]. In this way the background of γ radiation (e.g. synchrotron radiation) and internal noise (dark counts) can be suppressed very efficiently. In contrast to the analogue charge detection of most other BLM systems, coincidences are counted while the count rate is proportional to the loss rate as long as the number of overlapping coincidences is small. Counting of charged particles crossing both diodes has a few implications:

- Both channels need a discriminator to suppress dark counts due to noise. Since the signal of one minimum ionizing particle (MIP) is still weak, the threshold cuts also some of the MIP signals which reduce the efficiency. The efficiency for a coincident detection of MIPs was found to be about $\varepsilon_{count}$ = 30 % to 35 % per MIP including the readout electronic characteristic [171, 172].
- The dark count rate in coincidence mode is very small, typically <0.01 Hz
- The counter cannot distinguish between one or more MIPs crossing both diodes at the same time. The shortest signal length is defined by the response time of the diodes, but in practice it is defined by the readout electronics. An efficient counting type of BLM should have a signal length shorter than the bunch distance, so that the maximum measured loss rate is the bunch repetition rate of the accelerator. Saturation effects occur even before the maximum rate but they can be corrected by applying Poisson statistics [173].
- The dynamic range lies between the dark count rate of <0.01 Hz and the maximum rate (e.g. 10.4 MHz for HERA) and might reach $10^9$.

A PIN diode BLM system has been successfully operated between 1992 and 2007 in HERA without significant problems or radiation damage.

## 5.3 Secondary emission monitors

Gaseous ionization chambers have the disadvantage that their charge collection is slower than the bunch distance in most accelerators. Counting mode devices have to integrate the counts over a lot of bunches to get a statistically relevant signal. In some cases a bunch resolved fast signal is required, e.g. for fast machine protection [174]. A simple, robust and fast BLM is a secondary emission chamber. Secondary electrons are emitted from a surface due to the impact of charged particles with an efficiency of a few percent [175]. Secondary electron emission (SEE) is a very fast effect, but its very low sensitivity makes secondary electron emission useable only in high radiation fields, with the additional advantage that it consists of nothing more than a few layers of metal. Therefore, it is a very radiation-resistant monitor. The monitor has to be evacuated to avoid contamination of the signal due to gas ionization. Since the efficiency of gas ionization is much higher, a gas pressure of better than $10^{-4}$ mb should be achieved to get <1% signal from ionization. In particular, in high-radiation fields gas ionization will lead to non-linearities while secondary electron emission is a very linear process over a wide range of intensities [175, 176]. Unavoidable ionization at the feedthroughs and connectors limits the linearity at the lower end of the signal; the upper end is not seriously studied [151]. A dynamic range of $\gg 10^5$ is expectable [174].

A SEE multiplier extends the use of SEE BLMs to small radiation intensities. As long ago as 1971 aluminum cathode electron multipliers (ACEMs) have been used for beam loss measurements [177]. This device is a PMT where the photocathode is replaced by a simple aluminum cathode. The SEE electrons are guided to dynodes where they are amplified; amplifications up to $10^6$ are possible. An example for recent use of ACEMs can be found in Refs. [178, 179].

### 5.4 Scintillation detectors

SEE-based BLMs are very fast but still have a moderate sensitivity. An equivalent speed (a few nanoseconds) but much higher sensitivity can be achieved with scintillation counters: a combination of a scintillating material and a PMT. Large area plastic (organic) and liquid scintillators are available. In particular, plastic scintillators can be modulated in nearly all shapes and sizes while inorganic scintillators are expensive and limited in size. Descriptions of details of the scintillation process can be found in Ref. [180] and in various text books, e.g. Refs. [181, 182]. Large scintillators can be useful to enhance the solid angle of beam loss detection if the resulting radiation is not distributed uniformly. This is often true if the BLM is located very close to the beam pipe where the radiation is peaked into a solid angle and at low beam energies. Typically a thin layer of scintillator (0.3 cm to 3 cm) is sufficient to ensure sensitive loss detection, even at very limited space conditions [183].

Note that the light transmission through the scintillator (and the light guide) changes due to radiation damage. This depends strongly on the scintillator and light guide material, but for organic scintillators a typical value can be assumed: the transmission decreases to $1/e$ of its original value after about 0.01 MGy to 1 MGy (1 Mrad to 100 Mrad) collected dose. Liquid scintillators are somewhat radiation harder and have about the same sensitivity [184]. Inorganic scintillators such as BGO or CsJ(Tl) have about a factor of 10–50 higher sensitivity but their radiation resistance is poor and large size crystals are very expensive.

The gain of the same type of photomultipliers ($PMT_{gain}$) varies within a factor of 10. Therefore a careful inter-calibration of the BLM sensitivities is necessary by adjusting the high voltage (HV). The drift of the gain is a well-known behaviour of PMTs. A stabilized HV source and continuous monitoring of the photomultiplier gain over the run period are necessary to keep the calibration error small. The adjustable gain of the PMT increases the dynamic range of this type of BLM. At high gain the noise of a PMT is still quite low but non-linearities appear at low gain and high losses in the PMT; the space charge of the signal cloud cannot be compensated any more by the low voltage between the dynodes. A dynamic range of $10^8$ was measured at LEDA [185].

A special BLM uses Cherenkov light created in the glass tube of the PMT which is then detected directly [186]. It is a quite radiation-tolerant system; however, the darkening of the PMT glass has to be compensated for by increasing the PMT gain. Such a system is not sensitive enough to measure "small normal" losses but it is used to control and limit strong and dangerous losses.

Cherenkov light created in long optical fibres is used to determine the longitudinal position of beam losses. The fast response of the Cherenkov signal is detected with photomultipliers at the end of the irradiated fibres. A time measurement provides the position measurement along the fibre while the integrated light amplitude gives the amount of losses. A longitudinal position resolution of

20 cm (= 1 ns at $v = 0.66c$) is possible. High-purity quartz fibres (Suprasil) withstand $30 \times 10^9$ rad and generate no scintillation. Scintillating fibres are about 1000 times more sensitive but are not very radiation hard [187]. Examples for Cherenkov fibre-based BLM systems can be found in Refs. [188, 189].

So far all detectors are sensitive to "local" losses that occur within proximity of the detector. Hadron beam losses are typically connected with higher neutron flux, while neutrons can travel quite a long distance along the accelerator. Therefore, neutron detectors (NDs) are good at detecting losses occurring metres away from the detector itself. This makes NDs hard to interpret but more reliable for **Machine Protection System** (MPS) purposes. Solely relying on "normal" BLMs can lead to hiding of losses because a machine tuning process sometime moves the loss to a place where it is not seen by the "normal" BLM. An example is the SNS ND with a PMT + scintillator inside an X-ray shielding (lead) and surrounded by a polyethylene neutron moderator [190]. It is used in addition to ionization chambers and scintillator PMTs.

## 5.5 Summary

Different types of beam losses together with some examples have been shown. Beam loss monitoring techniques for measuring losses along an entire accelerator have been discussed with a focus on the sensitivity of the various types.

The most common BLM is a short ionization chamber. Whether a simple air-filled chamber is adequate or an argon- or nitrogen-filled chamber with superior higher dynamic range must be used depends on the conditions of the particular accelerator. Ionization chambers can be built very radiation resistant.

Long ionization chambers using a single coaxial cable work well for one-shot accelerators or transport lines. To achieve spatial resolution of losses along an entire accelerator either the first two or the third of the following conditions must be fulfilled: (1) the machine must be much longer than the bunch train; (2) the particles must be relativistic; (3) the long chamber has to be split into short parts which are readout individually.

PIN diodes with thick depletion layers can be used as "solid-state" ionization chambers. They have a high sensitivity but they exist only in small sizes. The combination of two PIN photodiodes in a coincidence counting mode results in a detector with very large dynamic range and extremely effective rejection of noise. A limitation is the inability to distinguish overlapping counts, so that the response is linear only for losses which are less than one count per coincidence interval.

A very fast and sensitive BLM system is a PMT in combination with a scintillator. Owing to the adjustable gain the dynamic range can be large, but the calibration of each device must be adjusted and monitored over time.

Long optical fibres can be used as in long ionization chambers with the same limitations in the bunch repetition rate. Cherenkov-based fibres are much more radiation hard but much less sensitive to losses than scintillating fibres.

Table 3 summarizes the different BLM types used in various high-intensity hadron accelerators.

**Table 3:** BLM types used at some high-intensity hadron accelerators

| | | |
|---|---|---|
| **Scintillator** | | |
| LEDA | CsI scintillator PMT based | [185] |
| ISIS | Plastic scintillator (BC408) | [183] |
| J-PARC RCS, MR, LINAC | GSO scintillator | [148] |
| SNS Ring | Scintillator PMTs | [190] |
| SNS Linac | PMTs with a neutron converter | [190] |
| PSR | Liquid scintillator with PMT (old) | [191] |
| CSNS | Scintillator PMTs | [192] |
| **Ionization chambers** | | |
| LEDA | 160 cm$^3$ N$_2$ ion chamber | [193] |
| ISIS | Long Ar ionization tubes (3 m to 4 m) | [183] |
| SNS Ring | 113 cm$^3$ Ar ion chambers | [152] |
| SNS Linac | 113 cm$^3$ Ar ion chambers | [190] |
| PSI | Air ionization chambers | [194] |
| PEFP | | |
| J-PARC RCS, MR, LINAC | Ar+CO$_2$ proportional counters (80 cm) and coaxial cable ion chambers, air filled (4 m to 5 m) | [148] |
| PSR | ion chambers filled with 160 cm$^3$ of N$_2$ gas | [191] |
| LANSCE | 180 cm$^3$ N$_2$ ion chamber | [195] |
| CSNS | 110 cm$^3$ Ar ion chamber | [196] |
| AGS | Ar-filled long coaxial ion chambers | [163, 197] |
| NuMI | Ar-filled Ion glass tubes | [156] |
| SPS, CNGS | Air-filled ion chambers (1 litre) | [192] |
| APT | Same as LEDA | |
| Tevatron, MI, Booster | Ar-filled Ion glass tubes, 190 cm$^3$ | [198, 199] |
| CERN LHC | N$_2$-filled ion chambers 1.5 litre | [142] |
| Rhic | Ar-filled Ion glass tubes | [197] |
| **SEM chambers** | | |
| LHC | SEM chambers | [175] |
| **PIN diodes** | | |
| HERA | PIN diodes in counting mode | [145] |
| Tevatron | PIN diodes in counting mode | [200] |
| Rhic | PIN diodes in counting mode | [201] |

We now give some examples for beam diagnostics for high-intensity hadron beams.

# 6 Transversal beam halo measurements

Particles which are expelled from the beam core form a halo around the beam. This halo can cause harmful beam losses, especially at higher beam energies. It contributes to activation of the environment and to background in the experiments. There are numerous sources of halo formation, in linear and circular accelerators, which are summarized in Ref. [202].

In the summary of the HALO'03 workshop [203] is written: "…it became clear that even at this workshop (HALO'03) a general definition of 'Beam Halo' could not be given, because of the very different requirements in different machines, and because of the differing perspectives of instrumentation specialists and accelerator physicists… From the diagnostics point of view, one thing is certainly clear – by definition halo is low density and therefore difficult to measure…". A quantification of the halo requires a more or less simultaneous measurement of the core and the halo of the beam. Therefore, halo measurements require very high dynamic range instruments and methods as well as very sensitive devices to measure the few particles in the halo. The difference between "halo" and "tail" can be defined as tails are deviants from the expected beam profile of the order of a few percent or per mille while halos are much less than this.

A measurement of the halo should result in a quantification of the halo; therefore, it is important to have a definition of the halo in at least one-dimensional spatial projection since this is relatively easy to obtain by a beam profile/halo monitor. For a complete understanding of the halo it might be necessary to extend the one-dimensional work to the whole phase space, in the measurement (location of the monitors) as well as in the theoretical work. This leads finally to the kinematic invariants imposed by Hamilton's equations [204]. Such a consideration is mainly used in simulations [205, 206].

In any case, the separation between the halo and the main core of the beam is not well defined. This leads to uncertainties to define a good description of the halo content of a beam. Typically beam halo is defined as an increased population of the outer part of the beam relative to the expected distribution which describes the core. Three different methods are commonly in use to characterize beam halo:

- kurtosis [204, 206, 207];
- Gaussian area ratio [208];
- ratio of beam core to offset [209].

An important feature of such quantifiers is that they are model independent and rely only on the characteristics of the beam distribution itself. Note that a measurement always contains instrumental effects. To define the halo contents in such a theoretical way one has to exclude these effects in advance.

The following sections concentrate on the instruments which are able to measure the beam halo and its evolution directly. Since the definition of halo is something like "< $10^{-4}$ of the beam core", some usual beam profile monitors might have intrinsic limitations to get the required dynamic range. For example, ionization-beam profile monitors (IPM), luminescence beam profile monitor and laser-based monitors are not (yet) sufficient for very high dynamic range halo measurements [210].

## 6.1 Beam halo measurements with wire scanners

Wire scanners are widely used for halo measurements with huge dynamic range and high sensitivity. This instrument provides a direct halo measurement by analysing the signal amplitude directly or in combination with particle counting. A combination of a wire and a scraper can be used to improve the sensitivity. Typically the signal is read out by the secondary electron emission current of the wire (low beam energy) or by scintillators measuring the scattered particles (high beam energy). The problems of wire scanners are well known, e.g. emittance blow-up and wire heating (see Section 4).

The direct beam profile and halo measurement is done by correlation of the signal with the position of the wire with a high dynamic range readout. A dynamic range of $10^5$ was achieved by linear amplification and the use of a 16-bit ADC as well as using a logarithmic amplifier which allows a standard 12-bit ADC [211]; see Fig. 17. The use of the secondary electron emission signal in low-energy accelerators has the advantage of avoiding an intrinsic error of measuring asymmetric tails by the asymmetric location of external detectors and/or large beam offsets [212]. This effect vanishes at higher energies and smaller beams. Therefore, scintillation counters outside the vacuum chamber can be used to measure the amount of scattered and shower particles created at the wire. Such scintillators are also sensitive to background due to, e.g., residual gas scattering, bremsstrahlung and other sources of beam losses. A telescope counter using a coincidence technique can reduce this background dramatically as well as dark counts (noise) from the counters itself so that a dynamic range of $10^8$ can be achieved [213–216]. The lower limit is defined by the remaining background rate.

In low-energy accelerators and/or at low bunch repetition rates as in a linac the counting method might not be very useful. In addition, a secondary electron emission current readout of a thin wire in the beam halo does not deliver enough current for a reliable measurement. Therefore, the wire size has to be increased even to a solid scraper to increase the achievable signal [217]. Their halo scanner consists of a 33 μm carbon fiber and two halo scrapers consisting of two graphite plates. Special care has to be taken that the beam does not induce too much heating of the scraper. Like in the counting method, the wire scanner and two scraper data sets must be joined to plot the complete beam distribution for each axis [218].

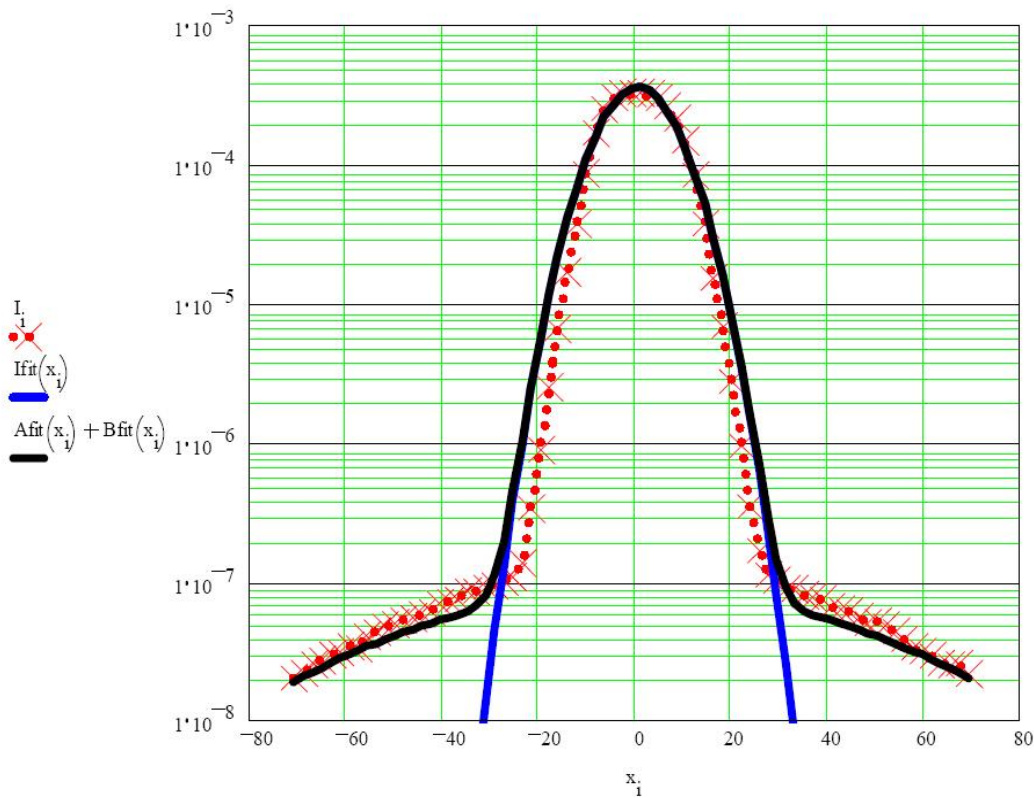

**Fig. 17:** A normal function shown in solid blue has been fit to the data (red crosses). A sum of two normal functions is shown in solid black. The *x*-axis is scaled as scanner position in millimetres and the *y*-axis is log-ampere input current in Amperes [211].

## 6.2 Beam halo measurements with interceptors/scrapers

Halo collimators are designed to remove the halo of the beam, but halo measurements can also be performed by moving one jaw of a collimator closer to the beam in steps. Either the beam current or the signal from adjacent BLMs can be recorded for each jaw position. The derivate of the signal gives the halo distribution. Very high sensitivity can be achieved by using BLMs close to the collimator jaws. The signal of the BLMs is proportional to the inverse lifetime of the beam which gives loss rates directly in terms of equivalent lifetimes. By moving the collimators closer until significant lifetime reductions were observed, the lifetimes calculated from beam currents can be used to calibrate the BLMs. Since this scraping method is a slow process it is very important to normalize each data point to the measured beam, to the measured beam size of the beam core and to the beam position [219].

Note that in high-energy and/or high-intensity accelerators/storage rings a complete scan of the whole beam is impossible since the jaws are typically not designed to withstand the full beam intensity [220, 221]. Therefore, a calibration of the halo contents (relative to the beam core) is often not possible or contains large errors, but relative changes of the halo can be detected at a very low level and far outside the beam core, e.g. ground motion frequencies and diffusion parameters [222–224].

Note also that in a synchrotron one jaw of a collimator will always scrape both sides of the beam distribution due to the β oscillation of the beam particles. Therefore, one will always measure a symmetric halo distribution.

Instead of a collimator with BLM readout other sensitive detectors can be moved into the halo to generate directly a signal from the halo particles. Various techniques are reported using, e.g., ionization chambers [225], scintillation fibres [226], vibrating wire scanners [227] or secondary electron emission foil [228]. All of these devices have the same strong limitation in determining the halo relative to the beam core, but relative changes of the halo can be observed with high sensitivity and resolution.

### 6.3 Optical halo monitors

For hadron beams optical methods are barely used since electromagnetic light generation (e.g. by synchrotron radiation, optical transition radiation) by hadrons is suppressed due to their high mass. Therefore, it is discussed here only very briefly.

The previously discussed methods to measure the halo distribution are relatively slow. Scanning of the halo typically needs seconds to minutes. One needs a stable beam and precise correlations with the beam size and position are mandatory. Optical methods have to give enough light to measure the core of even one single bunch at one passage. The light generation of these effects is linear over a huge dynamic range.[10] The dynamic range of the light detector (e.g. CCD cameras) can be improved by special optical systems:

- CID camera system with a dynamic range of $>10^8$ (see Ref. [229]);
- micro-mirror array [230].

Most optical applications suffer from diffraction limits which create diffraction fringes of $10^{-2}$ to $10^{-3}$ of the peak intensity which makes halo observations of lower than $10^{-3}$ impossible. A coronagraph with a so-called 'Lyot stop' [231, 232] removes this fringes and a background level of $6 \times 10^{-7}$ was observed. More details can be found in Ref. [210].

## 7 Longitudinal beam halo measurements

The meaning of "longitudinal halo" can be divided into three different classes of different interests:

- *Beam in the abort gap*. High-intensity and superconducting hadron storage rings need a gap in the bunch train to have enough time for loading the dump kickers to ensure a clean beam dump. In the case of a beam dump any particles in the gap will be lost around the ring risking a quench.
- *Coasting beam*. Experiments in colliders need very clear background conditions and precise time structures of the bunch crossings. Particles outside the main bunches may contribute to background as well as to undefined timing of the trigger counters in the experiment.
- *Neighbour bunches or bunch purity*. In time resolved experiments on synchrotron light sources a clear signal from one bunch without contributions from the adjacent (neighbour) buckets is desired. The neighbour bunches have to be determined on level of better than $10^{-6}$. This topic is mainly related to synchrotron light sources and will not be discussed here.

---

[10] The light generation in scintillation and phosphor screens suffers from non-linearities [233, 234] and therefore might not be applicable for huge dynamic range measurements.

## 7.1 Beam in gap

Stringent particle loss constraints in high current accelerators and in superconducting machines require a clean beam gap. Extraction of the beam (to the experiments or to the dump) is done by kickers with limited rise times, typically a few microseconds. This time is known as the abort gap where no particles should be stored. Any beam in this gap (bunched or coasting beam) will spray around the machine if the dump kicker is fired causing some problems:

- quenches (superconducting magnets);
- activation;
- spikes in experiments;
- equipment damage.

Reasons for beam in the abort gap can be:

- injection errors (timing);
- space-charge pushing particles out of the RF bucket;
- debunching;
- diffusion;
- RF noise/glitches;
- other technical problems.

Therefore, a continuous determination of the amount of beam in the gap is necessary to either clean the gap[11] or dump the whole beam before major problems arise. In high-energy storage rings like the LHC, Tevatron or HERA the presence of particles in the gap can be detected by the synchrotron radiation they emit, using the synchrotron radiation profile monitor port. Note that in principle any other fast process, e.g. beam-induced gas scintillation or secondary electron emission or BLM signals (e.g. at halo scrapers) can serve as a signal source, which are not limited to very high beam energy [233–235]. A fast and gateable detector which is synchronized by the revolution frequency is most useful to avoid saturation due to the signal of the main bunches. Optical methods have the advantage of existing detectors which are fast and sensitive enough to measure even a small amount of beam in the gap. A gated MCP PMT is able to measure both components of the beam in the gap, the bunched (AC) and the unbunched (DC) components while an intensified gated CCD or CID camera integrates over many turns and measures the DC component only [236]. Typical gate rise times of about 1 ns are sufficient for this application. Often the display of the analogue signal of the MCP PMT versus the gate time is sufficient but the dynamic range is limited to about $10^3$ due to the noise of the PMT. When using the gating technique one has to take into consideration the maximum duty cycle of the instrument. A typical maximum duty cycle of 1 % (e.g. Hamamatsu R5916U-50 MCP PMT) might not allow a complete gate over the whole gap at every turn. Therefore, the gate repetition rate has to slow down or a shorter gate has to be scanned across the gap [237]. The dynamic range and the signal-to-noise ratio can be increased by applying the time-correlated single photon counting method. First results with MCP PMTs and fast avalanche photo diodes are reported in Refs. [238, 239].

## 7.2 Coasting beam

The coasting beam is the part of the beam which is not captured by the RF system; its energy is not being replenished by the RF system. Even in high-energy hadron storage rings uncaptured protons lose only a few electronvolts per turn so that they can be stored for many minutes up to hours. Uncaptured beam slowly spirals inward and is lost on the tightest aperture in the ring. RF noise, glitches or intra-beam scattering can cause diffusion out of the RF buckets leading to coasting beam [240, 241]. The total uncaptured beam intensity is a product of the rate at which particles leak out of the buckets and the time required for them to be lost. This kind of beam loss causes additional activation of the collimators as well as additional background in the experiments. In particular, collider experiments might suffer from this background; therefore, they are interested in measuring the amount of coasting beam. Very sensitive methods are needed to measure small fractions of coasting beam in an appropriate time. Therefore, the experiments themselves use their sensitive detectors and fast trigger equipment which have a very large detection efficiency as well as very small dead times. Detailed measurements of coasting beam are reported from HERA-B (HERA) and CDF (Tevatron) [242–244]. Both detectors use as the signal source the beam loss in the detector while HERA-B even increased the loss rate using its internal wire target. The time structure of the losses is measured by fast counters and TDCs (HERA-B) or by integrating counts versus a sliding time interval (CDF).

---

[11] Abort gap cleaning by, e.g., fast kickers, resonant excitation, electron lens, etc. are not discussed here.

Note that the signal source comes from the far transverse halo of the beam. Its time structure might differ strongly from the time structure of the beam core [242], especially because uncaptured beam slowly spirals inward. Therefore, a total determination of the amount of coasting beam will have a large uncertainty. As soon as the amount of coasting beam is large enough an absolute determination can be done by comparing the AC and DC beam current monitors readings. The DC monitor measures all circulating particles while the AC monitor is sensitive only to the bunched beam component (Fig. 18). The calibration of both monitors to each other can be done just after finishing the acceleration where no coasting beam had survived.

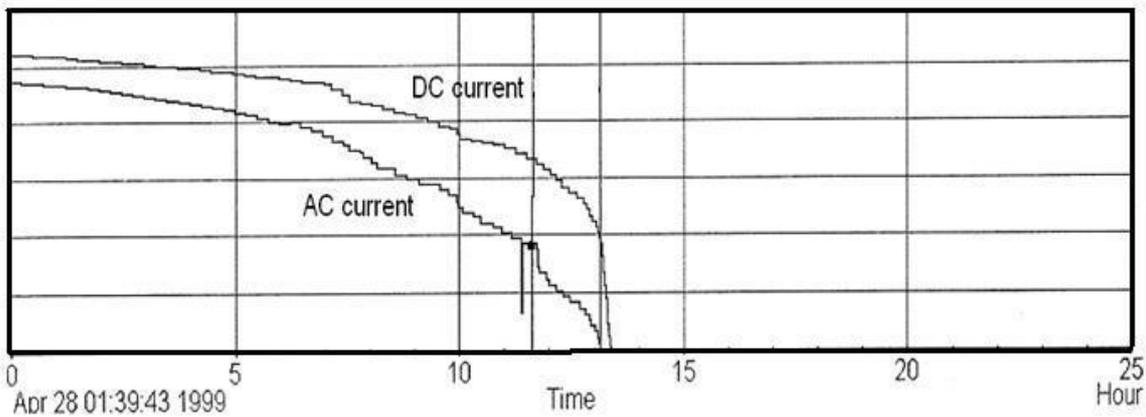

**Fig. 18:** DC beam current (includes coasting beam) and AC beam current (sum of all bunches) in HERA during an unusual store with a large amount of increasing coasting beam

## 8    Diagnostics for electron clouds

A charged beam can generate low-energy electrons by various, often unavoidable effects such as synchrotron radiation, residual gas ionization or stray particles. These electrons can strike the beam pipe wall and can create multiple electrons leading to multipactoring. Repetitive bunch crossings can lead to a quasi-stationary electron cloud (EC). A charged beam might interact destructively with this cloud resulting in beam instabilities and particle losses. Since the electrons are able to desorb gas from the wall, the first hint of an EC is typically an increase in the vacuum pressure in that section. The increase and observation of the vacuum is quite a slow process [245] and not very suitable for detailed analysis of ECs. Some more suitable instruments for EC diagnostics are discussed in the following sections.

### 8.1    Shielded BPMs

In front of the electrode of a button-type BPM, a grounded grid shields the electrode against the wake fields of the bunch. While the electrode is positive biased against ground it collects all low-energy electrons in its vicinity. A variable DC bias voltage enables electrons to be attracted or repelled depending on their energy. Such a relatively simple device is able to obtain time resolved information on the EC density (e.g. build up and decay) but an estimate of the EC line density λ is also possible:

$$\lambda = I_e /(e \cdot f_b \cdot tr \cdot A_e / A_{ch})$$

where $I_e = U_e/Z_e$ is the measured current on the electrode ($Z_e$ is the impedance of the electrode), $f_b$ is the bunch frequency, $tr$ is the transparency of the grid, $A_e$ is the area of the electrode and $A_{ch}$ is the inner area of the chamber.

In Refs. [245, 246] the resonant build-up of an EC is clearly diagnosed with such a shielded button-type BPM. Figure 19 shows a sketch of the design of a shielded pickup in the CESR-TA ring.

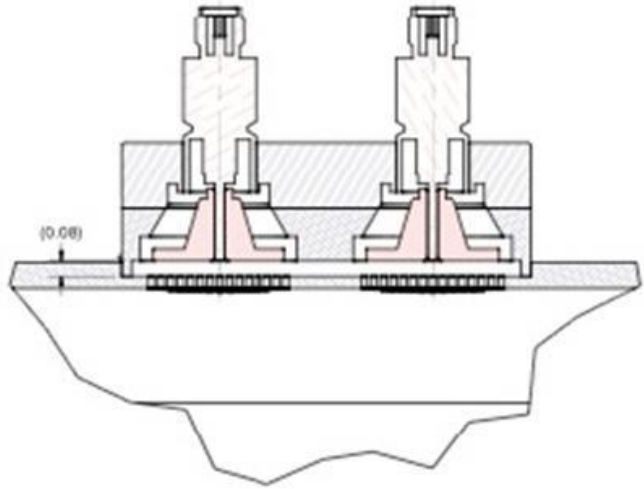

**Fig. 19**: Sketch of the CESR-TA shielded pick-up buttons [246]

## 8.2 Retarded field analyser

A retarded field analyser (RFA) is based on the same principle as the shielded BPM but it has a second retarding grid between the shield and the electrode [247]. The second grid is biased at a retarding potential ($E_r$) such that only electrons with kinetic energies greater than this are transmitted to the electrode (collector). The collector has a low secondary emission yield and is biased by a positive voltage. To amplify weak signals a MCP or channeltron can be used but usually the signal of ECs are sufficient for electronic amplification. The advantages of a dedicated RFA with respect to a shielded BPM are:

- increased surface area;
- higher sensitivity;
- better energy separation (see Fig. 20).

Examples of the extensive use of RFAs can be found in Refs. [248, 249].

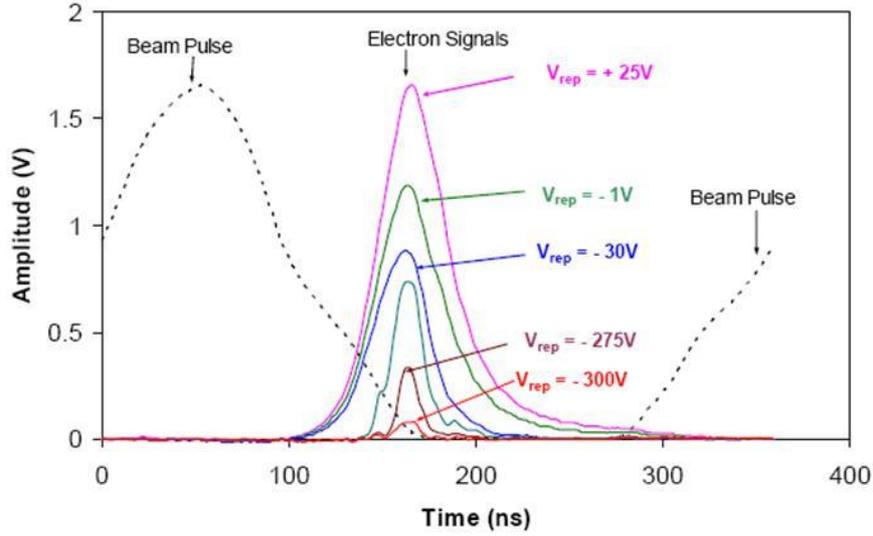

**Fig. 20:** RFA signals with different retarding voltage in time reference to the circulating beam pulse at PSR [248]. This experiment shows the low-energy distribution of the electrons of the cloud.

### 8.3 EC diagnostic with microwaves

The discussed EC monitors are suitable for localized measurements only. If the EC is created not in the vicinity it will not be detected by the monitor. In contrast, microwave transmission measurements are sensitive to the average EC density over a long section of the accelerator.

If a microwave of frequency ω is transmitted through electron plasma (EC) of length $L$ it will undergo a phase shift $\Delta\phi$:

$$\Delta\varphi \approx \frac{L \cdot \omega_p^2}{2c \cdot \sqrt{\omega^2 - \omega_c^2}} \quad \text{with} \quad \omega_p = \sqrt{\frac{N_e e^2}{\varepsilon_0 m_e}} \quad \text{the plasma frequency}$$

where $N_e$ is the electron density (typically $10^{11}$ m$^{-3}$ to $10^{12}$ m$^{-3}$ in ECs), $e$ is the electron charge, $\varepsilon_0$ is the vacuum permeability, $m_e$ is the electron mass and $\omega_c$ is the cutoff frequency of the beam pipe [250].

Therefore, the phase shift depends only on the electron density $N_e$ while all other parameters are constant and relatively well known.

The setup of a transmission measurement is shown in Fig. 21(a). For exciting a TE wave into the beam pipe a BPM can be used which has to be optimized for TE mode emission by using 180° hybrids and combiners. Splitting the power between pairs of opposite buttons, or striplines, lowers the power on a single electrode and improves the coupling to the TE mode electric field [251]; see Fig. 21(b). Note that the reversed power from the beam signal may disturb the signal generation. Hybrids are also used on the receiving BPM to suppress common beam position signals. The carrier frequency ω is chosen by measuring the optimum of the transmission function (obviously above $\omega_c$). With a constant EC a phase shift could hardly be detected. Therefore, one has to change the EC density during the measurement, typically by having a long enough gap between bunch trains to remove the EC. This gap creates than a phase modulation at the receiver (e.g. spectrum analyser) which appears as side bands to the carrier frequency in a distance of the revolution frequency. Its amplitude relative to the carrier is proportional to the phase shift $\Delta\phi$.

DeSantis, at ECLOUD'10, stated "Although having a simple formulation, the practical application of the TE wave method is not straight forward". Many problems might hinder the analysis [251–253]:

- The coupling efficiency of BPMs is small above cutoff, impedance is not well matched (by design).
- Non-linearities by reflections in generator and receiver add sidebands to the spectrum.
- The strong beam harmonics superimpose the weak EC sidebands.
- AM modulation by resonant coupling to $e^-$ trapped in the magnetic field (near the cyclotron frequency) add sidebands.
- Owing to reflections of the carrier, $L$ can be underestimated.
- $L$ is not always the distance between the BPMs, the cloud might be shorter.
- Ensure that cleaning gap is larger than the decay time of cloud. Take into account EC rise and fall times.
- How precise the cutoff frequency $\omega_c$ is known?

A local transmission measurement (below cutoff) can be done with the setup of Fig. 21(c). First tests were done in Ref. [253] but a complete understanding of the physics of this method is still under discussion.

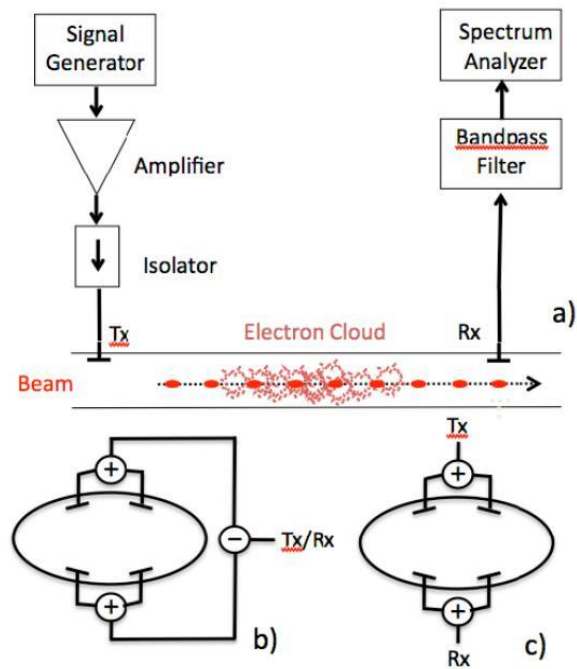

**Fig. 21:** (a) Microwave transmission setup (Tx, transmitter; Rx, receiver), (b) BPM arrangement for transmission measurement (⊕, splitters, ⊖, 180° hybrid), (c) BPM arrangement for "local transmission" measurement (reproduced from [253])

## 9  Injection mismatch

As a rule, proton/ion accelerators need their full aperture at injection, thus avoiding mismatch allows a beam of larger normalized emittance $\varepsilon_n$ and containing more protons. In proton/ion ring accelerators any type of injection mismatch will lead to an emittance blow-up. Off-axis injection will lead to orbit oscillations. These oscillations can be detected easily by turn-by-turn BPMs in the ring (before Landau damping occurs). The orbit mismatch can be corrected by a proper setup of the steering magnets, kickers and septa. Any mismatch of the optical parameters α, β, γ (space charge), however, will also lead to an emittance blow-up (and beam losses) and is not detectable by BPMs.

Figure 22(a) shows the phase ellipse at a certain location in a circular accelerator. The ellipse is defined by the optics of the accelerator with the emittance ε and the optical parameters β (beta function), $\gamma = (1 + \alpha)/\beta$ and the slope of the beta function $\alpha = -\beta'/2$. Figure 22(b)–(d) show the process of filamentation after some turns.

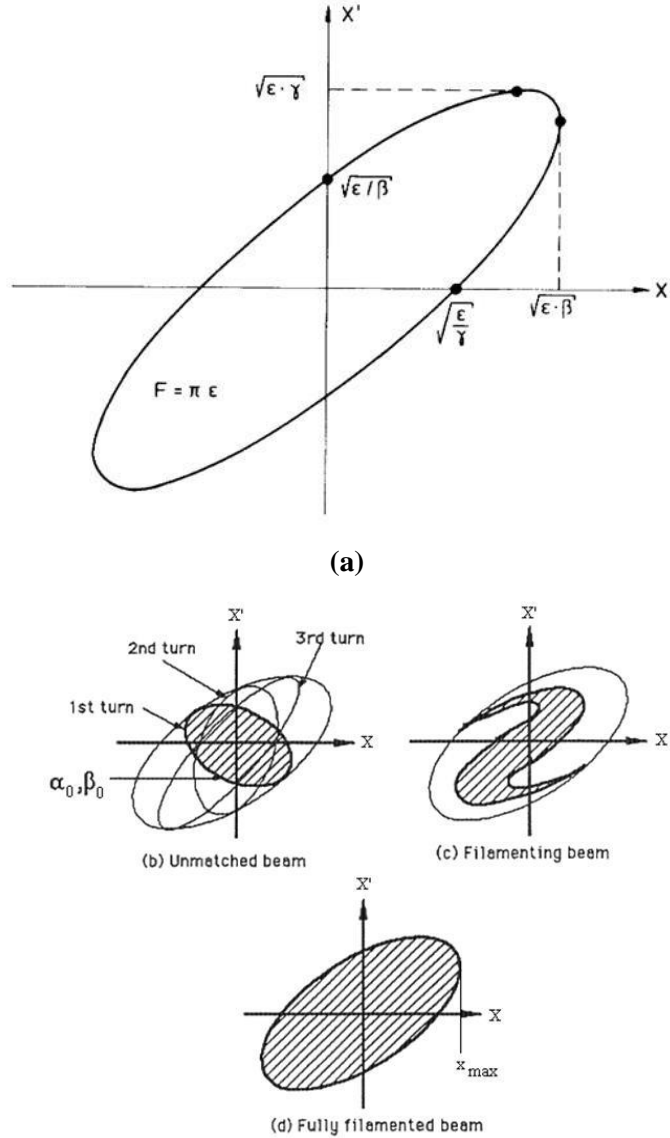

**Fig. 22:** (a) A phase space ellipse of a circular accelerator, defined by α, β, γ, ε. (b) Filamentation of an unmatched beam. (Reproduced from Ref. [254].)

Assuming a beam is injected into the circular machine, defined by $\beta_0$ and $\alpha_0$ (and, therefore, $\gamma_0$) with a given emittance $\varepsilon_0$. For each turn $i$ in the machine the three optical parameters will be transformed by

$$\begin{pmatrix} \beta_{i+1} \\ \alpha_{i+1} \\ \gamma_{i+1} \end{pmatrix} = \begin{pmatrix} C^2 & -2SC & S^2 \\ -CC' & SC'+S'C & -SS' \\ C'^2 & -2S'C' & S'^2 \end{pmatrix} \cdot \begin{pmatrix} \beta_i \\ \alpha_i \\ \gamma_i \end{pmatrix} \quad \text{(starting with } i=0\text{)}$$

where $C$ and $S$ are the elements of the Twiss matrix ($\mu = 2\cdot\pi\cdot q$, where $q$ is tune):

$$\begin{pmatrix} C & S \\ C' & S' \end{pmatrix} = \begin{pmatrix} \cos\mu + \alpha_0 \cdot \sin\mu & \beta_0 \cdot \sin\mu \\ -\gamma_0 \cdot \sin\mu & \cos\mu - \alpha_0 \cdot \sin\mu \end{pmatrix}$$

and $\gamma = (1 + \alpha^2)/\beta$

Without any mismatch, the three parameters will be constant while a mismatch will result in an oscillation of the parameters at twice the betatron tune [255, 257]. A mismatch of, e.g., the betatron phase space will result in transverse shape oscillations, at least for some 10 turns, before the different phases of the protons lead to a filamentation of the beam. A measurement of width (or shape) oscillations at injection is a very efficient method to detect an optical mismatch that increases the emittance in the circular accelerator. A measurement of the turn-by-turn shape oscillation is possible with a fast (turn-by-turn) readout of:
(1) thin screen (OTR, phosphor); see Ref. [258] for details;
(2) secondary electron emission grids [259];
(3) IPM [260];
(4) quadrupole (QP) pickup [261];
(5) synchrotron radiation (SR) monitor (electrons) [262].

The effect of the monitors on the beam include the following:

- Screen/grid: emittance blow-up and losses.
- IPM: very small, a sufficient signal at each turn needs a pressure bump leading to emittance blow-up and losses.
- QP pickup: none but very difficult to suppress the dipole mode.
- SR monitor: none, but no light from protons at low energy.

## 9.1 Blow-up

A screen/grid or IPM pressure bump will give an additional constant increase of the emittance, but it can easily be separated from the oscillation observation. The protons receive a mean kick at each traverse through a screen resulting in an additional angle θ:

$$\theta = \frac{0.014}{p \cdot \beta} \cdot Z \cdot \sqrt{\frac{d}{l_{rad}}} \left[ 1 + \frac{1}{9} \log_{10}\left(\frac{d}{l_{rad}}\right) \right] \quad \text{in radians}$$

where $p$ is the momentum in GeV/$c$ and $Z = 1$ the charge number of the proton, $\beta = v/c$ the velocity, $d$ the thickness of the foil and $l_{rad}$ the radiation length of the material of the foil. This formula describes the Gaussian approximation of the mean scattering angle of the protons after one traverse. The change of the emittance $\delta\varepsilon$ for every turn can be calculated by

$$\delta\varepsilon_{rms} = \sqrt{2 \cdot \pi} \cdot \theta^2 \cdot \beta$$

which adds quadratic to the 1σ emittance of the previous turn. The emittance blow-up due to a thin foil is much too large at low energies. A harp of thin wires produces less emittance blow-up. Assuming a harp of 20 μm titanium wires at a separation of 1 mm, the blow-up can be calculated as in a 0.2 μm foil. The secondary electron emission current created in the wires can be read out by fast ADCs turn by turn. Such a readout schema is applied in the PS-Booster at CERN [258].

## 9.2 Losses

The relative proton losses per turn $dN/N_0$ in the foil (thickness $d$) is given by the nuclear interaction length $L_{nuc}$:

$$\frac{dN}{N_0} = \frac{d}{L_{nuc}} \quad \text{with} \quad L_{nuc} = \frac{A}{\rho \cdot N_A \cdot \sigma_{nuc}}$$

here $L_{nuc}$ depends on the total nuclear cross-section of the nuclear interaction $\sigma_{nuc}$, the density $\rho$ of the foil and the Avogadro constant $N_A = 6.0225 \times 10^{23}$ mol$^{-1}$. The nuclear cross-section $\sigma_{nuc}$ depends on the proton momentum and on the material of the foil and is shown for different materials in Table 4 between a momentum of $0.3 < p < 40$ GeV/$c$.

**Table 4:** Nuclear total cross-sections, interaction length and particle losses

| Material<br>$A$ [g/mol]<br>$\rho$ [g/cm³] | Momentum [GeV/$c$] | $\sigma_{nuc}$ [mb] | $L_{nuc}$ [cm] | Relative loss/turn $dN/N_0 \times 100$ [%] with $d = 10$ μm |
|---|---|---|---|---|
| Carbon | 0.3 | 280 | 31.5 | $3 \times 10^{-3}$ |
| 12.01 | 7.5 | 360 | 24.5 | $4 \times 10^{-3}$ |
| 2.26 | 40 | 330 | 22.5 | $4.4 \times 10^{-3}$ |
| Aluminum | 0.3 | 550 | 30.2 | $3.3 \times 10^{-3}$ |
| 26.98 | 7.5 | 700 | 38.4 | $2.6 \times 10^{-3}$ |
| 2.70 | 40 | 640 | 35.1 | $2.8 \times 10^{-3}$ |
| Copper | 0.3 | 950 | 12.4 | $8.1 \times 10^{-3}$ |
| 63.546 | 7.5 | 1350 | 17.6 | $5.7 \times 10^{-3}$ |
| 8.96 | 40 | 1260 | 16.4 | $6.1 \times 10^{-3}$ |

The loss rate is negligibly small even at the injection energies of proton machines and will not influence the mismatch measurement.

### 9.3 Some notes on the readout

The optical readout of screens/IPMs is slow. A turn-by-turn observation needs a 100 kHz (3 km) data collection of the whole image. Line sensors with a larger pixel size (for better sensitivity) nowadays have a readout frequency of >15 MHz/pixel. Assuming 128 pixels will give a maximum readout frequency of 117 kHz for a one-dimensional image.

A secondary electron emission signal as well as the QP pickup signal can be picked up with very high frequencies, even bunch by bunch (100 MHz) and is therefore preferred for smaller ring diameters with a higher revolution frequency.

## 10 Beam energy

### 10.1 Beam energy determination using spectrometers

The most common method for determining the momentum/energy of a particle is a spectrometer. This includes any circular accelerator where the main dipole field and the closed orbit, resp. the central frequency define the particle energy [263] while spectrometer magnets making use of this effect are widely used in hadron Linacs. Relative energy resolutions are of the order of $10^{-4}$ [264].

Spectrometers measure the particle momentum by precisely determining the angle of deflection Θ in a dipole magnetic field B:

$$\Theta \propto \frac{1}{p} \int B ds$$

A very good determination of the magnetic field ($10^{-5}$ or better) and the beam position at the entrance and exit of the spectrometer magnet is essential for a precise measurement. A position-sensitive detector at the end of the spectrometer arm enables a precise momentum and momentum spread measurement. A collimator in front of the spectrometer magnet and a detector position at a low β and high dispersion value improves the precision of the measurement [265].

## 10.2 Beam energy determination using TOF

The resulting profile at a spectrometer detector is a mixture of the transverse and longitudinal beam parameters. An independent measurement can be performed for non-relativistic energies using the TOF method.

Two or more fast beam pick-ups are installed in a straight section with a typical distance $L$ of several meters while $L$ has to be known exactly, with a typical precision of about 1 mm. Each kind of fast pick-up can be used as a signal generator; their well-known signal properties will define the start and end of the measured time $t$, e.g. maximum or half height of a unipolar signal, zero crossing of a bipolar signal (more precise). When picking up the same bunch its velocity β is simply given by $t = L/\beta c$, but the value has to be corrected for signal propagation delays along the cables [266]:

$$t_j = \frac{l_{cab,j}}{v_{cab}} + \frac{L_j}{\beta \cdot c}$$

where $v_{cab}$ is the cable phase velocity, $l_{cab,j}$ is the length of the cable of each station $j$ and $L_1 = 0$. Modern digital processing allows an I/Q method in a FPGA which results in better precision of the TOF measurement [267] as well as a possible comparison of the bunch with the cavity phase [268].

## 10.3 Energy measurement with other methods

The use of Rutherford scattering to extract the beam energy is limited to low energies only. In Ref. [269], a 0.3 mg/cm$^2$ thick gold foil was inserted into the beam periphery and the scattered protons were detected by two 500 µm thick silicon particle detectors. The detectors were placed at a distance of approximately 30 cm from the target, at angles of 45° and 100° with respect to the incident beam direction. Careful positioning of the foil in the beam halo is necessary to avoid saturation of the detector. The (full absorbing) detector measures the energy spectrum of the scattered particles with a strong peak at the beam energy. A fast detector (e.g. diamond) enables also a bunch length method with this technique [270].

It is possible to measure the energy of a laser- or gas-stripped electron of a H$^-$ beam. Beam electrons have the same velocity as the beam and therefore an energy of 1/1836 of the beam protons. A 200 MeV H$^-$ beam yields 109 keV electrons. The beam energy spectrum can then be determined by measuring the electron charge versus repelled voltage on a FC [271].

In Ref. [272] a longitudinal movement of a Feschenko-type monitor was proposed while the bunch shape functions are measured along a phase axis φ. Measuring Δφ and $d$ one can find the beam velocity β.

## 11 Machine protection systems

For this quite large topic I would like to refer to the comprehensive report of R. Schmidt on "Machine Protection" at CAS 2008 in Dourdan, France. Most of the pictures of the "Little Shop of Horrors" from the talk can be found in the recent ICFA Advanced Beam Dynamics Workshops "High Intensity High Brightness Hadron Beams".

## 12  Tune and chromaticity

During the presentations within this CAS a question about tune and chromaticity measurement was asked. The answer was given in the diagnostic talk "on the fly" and was not prepared as a special topic of high-intensity diagnostics in this report. Therefore, I would like to refer to the 5th workshop in the framework of CARE-N3-HHH-ABI , Novel Methods for Accelerator Beam Instrumentation, "Schottky, Tune and Chromaticity Diagnostic (with real time feedback)", 11–13 December 2007 in Hotel Prieuré, 74404 Chamonix Mont-Blanc, France, for tutorials and details about this diagnostic as well as for Schottky diagnostics [273].